\documentclass[a4paper,fleqn]{cas-dc}
\usepackage[numbers, sort&compress]{natbib}
\PassOptionsToPackage{hyphens}{url}
\usepackage{breakurl}
\usepackage{graphicx}
\usepackage{amsmath}
\usepackage{algorithm}
\usepackage{algorithmic}
\usepackage{enumitem}

\makeatletter
\def\UrlAlphabet{%
      \do\a\do\b\do\c\do\d\do\e\do\f\do\g\do\h\do\i\do\j%
      \do\k\do\l\do\m\do\n\do\o\do\p\do\q\do\r\do\s\do\t%
      \do\u\do\v\do\w\do\x\do\y\do\z\do\A\do\B\do\C\do\D%
      \do\E\do\F\do\G\do\H\do\I\do\J\do\K\do\L\do\M\do\N%
      \do\O\do\P\do\Q\do\R\do\S\do\T\do\U\do\V\do\W\do\X%
      \do\Y\do\Z}
\def\UrlDigits{\do\1\do\2\do\3\do\4\do\5\do\6\do\7\do\8\do\9\do\0}
\g@addto@macro{\UrlBreaks}{\UrlOrds}
\g@addto@macro{\UrlBreaks}{\UrlAlphabet}
\g@addto@macro{\UrlBreaks}{\UrlDigits}
\makeatother

\def\tsc#1{\csdef{#1}{\textsc{\lowercase{#1}}\xspace}}
\tsc{WGM}
\tsc{QE}
\tsc{EP}
\tsc{PMS}
\tsc{BEC}
\tsc{DE}

\usepackage{float}

\begin{document}
\let\WriteBookmarks\relax
\def\floatpagepagefraction{1}
\def\textpagefraction{.001}

\shorttitle{CCQNet to long-tailed bearing fault diagnosis}

\shortauthors{Wei-En Yu et~al.}

\title [mode = title]{A class-weighted supervised contrastive learning long-tailed bearing fault diagnosis approach using quadratic neural network}



%
\author[1]{Wei-En Yu}[type=editor,
                        orcid=]





\credit{Writing - Original Draft, Methodology, Software, Data curation}

\affiliation[1]{organization={School of Instrumentation Science and Engineering, Harbin Institute of Technology},
    city={Harbin},
    country={China}}

\affiliation[2]{organization={Department of Industrial and Systems Engineering, The Hong Kong Polytechnic University},
    city={Hong Kong},
    country={Special Administrative Region of China}}

\author%
[1]
{Jinwei Sun}

\credit{Project administration, Conceptualization, Writing - Review \& Editing}


\author%
[1]
{Shiping Zhang}[type=editor,]
\cormark[1]
\credit{Funding acquisition, Project administration, Supervision}
\cortext[cor1]{Co-corresponding authors}
\ead{spzhang@hit.edu.cn}

\author%
[2]
{Xiaoge Zhang}[type=editor,
                ]
\cormark[1]
\credit{Funding acquisition, Project administration, Supervision}
\ead{xiaoge.zhang@polyu.edu.hk}

\author%
[1,2]
{Jing-Xiao Liao}[type=editor,]
\cormark[1]
\credit{Funding acquisition, Project administration, Supervision}
\ead{jingxiaoliao@hit.edu.cn}

\begin{abstract}
Deep learning has achieved remarkable success in bearing fault diagnosis. However, its performance oftentimes deteriorates when dealing with highly imbalanced or long-tailed data, while such cases are prevalent in industrial settings because fault is a rare event that occurs with an extremely low probability. Conventional data augmentation methods face fundamental limitations due to the scarcity of samples pertaining to the minority class. In this paper, we propose a supervised contrastive learning approach with a class-aware loss function to enhance the feature extraction capability of neural networks for fault diagnosis. The developed class-weighted contrastive learning quadratic network (CCQNet) consists of a quadratic convolutional residual network backbone, a contrastive learning branch utilizing a class-weighted contrastive loss, and a classifier branch employing logit-adjusted cross-entropy loss. By utilizing class-weighted contrastive loss and logit-adjusted cross-entropy loss, our approach encourages equidistant representation of class features, thereby inducing equal attention on all the classes. We further analyze the superior feature extraction ability of quadratic network by establishing the connection between quadratic neurons and autocorrelation in signal processing. Experimental results on public and proprietary datasets are used to validate the effectiveness of CCQNet, and computational results reveal that CCQNet outperforms  SOTA methods in handling extremely imbalanced data substantially.
\end{abstract}



\begin{keywords}
Bearing fault diagnosis, Supervised contrastive learning, Long-tailed distribution, Class-weighted loss function, Quadratic neural network.
\end{keywords}

\maketitle

\section{Introduction}
The health of bearings is critical to ensure the sound performance of rotation equipment commonly used in a broad spectrum of industrial applications~\cite{wang2020multi,cao2021novel}. The failure and malfunction of bearings, which accounts for 40\%-70\% of engine failures, results in serious economic loss and even lead to casualties~\cite{vencl2014diesel,islam2019reliable}. Hence, it is of paramount importance to maintain bearings in a serviceable condition. Towards this goal, accurate and timely bearing fault diagnosis is an essential measure to reduce the downtime, diminish repair costs, extend bearing life, and improve the reliability as well as operational safety of rotating machinery~\cite{ali2015accurate}. A variety of studies in the literature find that bearing failures oftentimes get manifested as vibration signals on the surface, and exhibit specific fault frequencies corresponding to different fault modes~\cite{smith2015rolling, randall2011rolling}. Considering the valuable information embodied in the vibration signals, a series of approaches have exploited it for condition monitoring and fault diagnosis of bearings, and the physical mechanism between vibration signals with corresponding fault types has been revealed by a variety of analytical tools, such as envelope analysis~\cite{randall2001relationship, mcfadden1984model}, time-frequency analysis~\cite{feng2013recent, stankovic1994method, yu2019concentrated}, and cyclostationarity analysis \cite{napolitano2016cyclostationarity}. 

In general, bearing fault diagnosis can be classified
into two categories: signal processing-based methods and data-driven methods. Signal processing-based methods utilize signal processing techniques, such as wavelet transform \cite{lou2004bearing}, short-time Fourier transform \cite{gao2015feature}, blind deconvolution, and adaptive mode decomposition \cite{fengAdaptiveModeDecomposition2017}, to manually extract relevant characteristics from the raw vibration signal to facilitate the determination of the specific fault type. Although these methods have a sound mathematical foundation, they require to tailor the usage of signal processing techniques and the tuning of model parameters for each specific scenario. Data-driven methods have the potential to overcome these shortcomings. Among data-driven methods, one representative approach is deep learning, particularly convolutional neural network (CNN), it can learn feature representations from the raw data to support fault diagnosis in an end-to-end fashion~\cite{azar2022semi,nemani2023uncertainty}. Additionally, CNNs possess the ability to handle large-scale data in high dimensional space, and it thus emerges as a prevailing choice for many applications. As a result, CNNs have also been extensively utilized in bearing fault diagnosis. For example, \citet{zhang2017new} proposed a Deep Convolutional Neural Networks with Wide First-layer Kernels (WDCNN) to process 1D vibration signal in an end-to-end fashion for bearing fault diagnosis.
Other improvements have been made to the CNN-based diagnosis model, such as Deep Residual Shrinkage Networks (DRSN) \cite{zhao2019deep}, Dislocated Time Series CNN (DTS-CNN) \cite{liu2016dislocated}, and probabilistic spiking response model (PSRM) \cite{zuo2022multi}. These refinements lead to a superior performance of CNN for fault diagnosis in different contexts, such as heavy noise, varying rotation speed, and cross-domain characteristics.

Despite the aforementioned achievements in CNN-based bearing fault diagnosis, most CNN-based methods rely on a balanced data to develop a good-performing model. However, in practical operating conditions, machines usually operate in a normal or healthy state for the majority of time while various types of failures only occur with quite a low probability~\cite{yang2019intelligent, chen2022imbalance}. 
Take nuclear power plants as an example, regular maintenance is conducted on bearings of rotation equipment, such as seawater booster pumps, to ensure their stability and prevent breakdowns. These maintenance activities significantly improve the reliability of nuclear power plants, the probability of abnormal events thus drops down to a low level~\cite{10158933,harunuzzaman1996optimization}. Statistically, the data for fault diagnosis commonly exhibits a long-tailed distribution, where the number of samples in healthy states significantly outnumbers those in the fault states. In the machine learning community, the healthy class is considered as the majority (head) class, while the fault classes are considered as the minority (tail) classes. Under such circumstances, the long-tailed dataset poses a significant challenge to the training of CNN. If the long tail feature existing in the data is neglected, the trained CNN model typically exhibits a poor performance manifested in the form of high bias and prone to overfitting. These factors, individually and collectively, eventually get translated as a considerable increase in the misclassification rate~\cite{zhang2021deep}.

To combat class imbalance, a typical strategy is to balance the distribution of data through resampling to augment the minority class. Resampling methods, such as oversampling \cite{yang2020improved} and undersampling \cite{khoshgoftaar2009feature}, are commonly employed to remediate the class imbalance. In addition, data generation methods, like GAN \cite{liu2019generative} and VAE \cite{li2021unified} have been developed to generate synthetic samples to enrich the minority class. However, these methods face fundamental limitations when dealing with extremely imbalanced data. Some studies indicated that these conventional methods experienced significant performance degradation (roughly 60\% accuracy) when the imbalanced rate exceeded 20:1 \cite{zhang2022class, peng2022progressively, hou2022contrastive}. In practical scenarios, it is common to encounter such extreme class imbalance with thousands of healthy data but only a few dozen instances of fault data.
Specifically, resampling methods might discard a substantial number of samples in the majority class and potentially result in the loss of valuable information retained in these discarded samples. Furthermore, the high repeatability of 1D signals makes it hard to acquire effective samples for training~\cite{chen2022imbalance}. On the other hand, synthetic data generation methods face challenges as it is unable to guarantee the quality of synthetically generated samples particularly when the training data is scarce. As reported in Ref.~\cite{zhang2022class}, synthetic data generation risks distorting the distribution of the actual data and causing overfitting problems.


To overcome these challenges, contrastive learning offers a new perspective as it targets to optimize a contrastive loss function to increase the separability between positive and negative samples, thus learning a more discriminative feature space~\cite{khosla2020supervised}. Contrastive learning has been proven to be highly effective for fault diagnosis, especially in the cases with a limited number of training samples \cite{10042974,10018491}. However, existing contrastive learning approaches are not yet tailored to address the long-tailed issue, and their effectiveness is unavoidably influenced by the class imbalance. As such, some studies combined contrastive learning with undersampling techniques to fight against the imbalanced dataset. For instance, \citet{zhang2022class} combined contrastive learning with undersampling to achieve a balanced data distribution and trained a machine learning model using contrastive learning to achieve satisfactory outcomes in the long-tailed bearing fault diagnosis.
Nevertheless, undersampling unavoidably causes the repetition of a small amount of samples in the tail and gives rise to limited representativeness in the generated samples and poor model robustness.


In this paper, we are motivated to tackle these issues by refining the contrastive learning approach from two perspectives. In the first place, we propose to adopt a polynomial neural network, and more specifically, the quadratic network, to enhance feature extraction capability in CNN. In essence, the quadratic network replaces the linear neurons with nonlinear quadratic neurons in the neural network, and such a replacement injects a heightened expressive power compared to the conventional first-order neural network according to approximation theory~\cite{fan2023one}. In addition, previous study has demonstrated the efficacy and interpretability of the quadratic network in bearing fault diagnosis when faced with strong noise \cite{liao2023attention}. Despite these encouraging findings, the central issue \textit{why quadratic network outperforms the conventional network when processing vibration signals} remains a mystery. In this paper, we also answer this question through a rigorous mathematical deduction and conclude that quadratic neurons are able to achieve local autocorrelation, thereby facilitating the extraction of fault-related features. Such finding is crucial for a thorough understanding of the decision-making mechanism in the quadratic network for fault diagnosis.

Secondly, supervised contrastive learning has been shown to guide the model to collapse to the vertices of a regular simplex on a hypersphere when dealing with balanced datasets \cite{graf2021dissecting}. However, imbalanced datasets have an explicit impact on the distribution of vertices when the model collapses, subsequently influencing the separability of features on the hypersphere. Considering the effectiveness of reweighting techniques in dealing with imbalanced data, such as focal loss \cite{wu2021learning} and cross-entropy loss \cite{han2022end}, we argue that they can be applied to contrastive learning, particularly supervised contrastive learning, to address the long-tailed problem in the distribution. Compared to the state-of-the-art literature, the contributions of this paper are summarized as below:
\begin{enumerate}
    \item We propose a class-weighted contrastive learning quadratic network (CCQNet) for long-tailed bearing faults diagnosis. We employ a quadratic network as a feature extraction backbone and combine it with class-weighted contrastive loss and logit-adjusted cross-entropy loss functions. {Our method achieves an improvement in the model's ability to handle imbalanced data via a powerful feature extractor and a re-balanced loss function.}
    
    \item Mathematically, we demystify the superior signal feature representation ability of quadratic networks by deducing and establishing the connection between autocorrelation and quadratic neurons. To the best of our knowledge, this is the first theory to explain quadratic networks from the perspective of signal processing.

    \item We conduct comprehensive experiments using the public dataset and our own dataset. Experimental results suggest that CCQNet outperforms other state-of-the-art methods, especially in extremely imbalanced data.
\end{enumerate}

The rest of the paper is structured as follows. Section 2 gives a brief review on supervised contrastive learning. Next, the proposed CCQNet and its main operators are explained in Section 3. In Section 4, several experiments are performed to verify the effectiveness of the proposed method. Finally, we conclude this paper in Section 5.

\section{Supervised contrastive learning}
The key idea of contrastive learning is to bring samples of the same label (positive samples) closer together and push samples of different labels (negative samples) to fall apart. The primary difference between self-supervised contrastive learning and fully supervised contrastive learning lies in how to select positive and negative samples. As shown in Figure \ref{fig:SCL}, the fully-supervised contrastive learning regards all the samples from the same class as positive samples and treats samples from all the other classes as negative samples to fully take advantage of labeling information \cite{khosla2020supervised}. In general, supervised contrastive learning consists of three steps:

\begin{figure}[pos=htbp]
    \centering
    \includegraphics{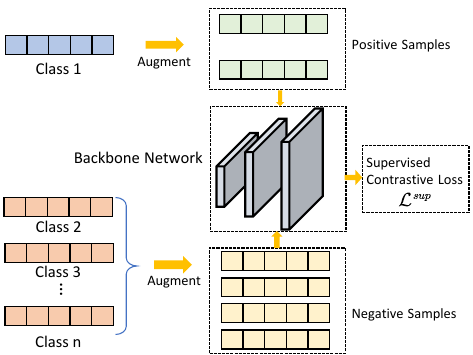}
    \caption{The framework of supervised contrastive learning.}
    \label{fig:SCL}
\end{figure}
(1) \textbf{Data augmentation}: Given an input dataset $\mathcal{S}=\{\boldsymbol{X} \in \mathbb{R}^{N\times n}; \boldsymbol{Y} \in \mathbb{R}^{N}\}$ consisting of sample/label pair $\{\boldsymbol{x}_i, y_i\}^{N}_{i=1}$, {two data augmentation methods, i.e., adding Gaussian noise, random scaling, are performed such that, each $\boldsymbol{x}_i$ is extended to two additional samples $\tilde{\boldsymbol{x}}_{i1}, \tilde{\boldsymbol{x}}_{i2}$, then raw samples are augmented as $\left\{\tilde{\boldsymbol{x}}_{i}\right\}_{i=1}^{2N}$, with $i \in \boldsymbol{I} \equiv \{1,2,\cdots,2N\}$.}

(2) \textbf{CNN construction}. The augmented samples are then passed through a backbone convolutional neural network $\text{CNN}(\cdot)$, followed by a projection head, such as a multi-layer perceptron $\text{MLP}(\cdot)$, and normalized to the vector $\boldsymbol{z}_{i}$:
\begin{equation}
\begin{aligned}
        \boldsymbol{u}_{i} &= \text{MLP}(\text{CNN}(\tilde{\boldsymbol{x}}_{i})),\\
        \boldsymbol{z}_{i} &= \frac{\boldsymbol{u}_{i}}{||\boldsymbol{u}_{i}||}.
\end{aligned}
\label{eq:mapping}
\end{equation}
where $||\cdot||$ denotes the $L_2$ norm.

(3) \textbf{Building contrastive loss function}. Assuming the mini-batch size is $B$, and the mapping network $\phi$ projects augmented data to their latent embeddings with $\phi: \tilde{\boldsymbol{X}}=\{\tilde{\boldsymbol{x}}_i\}_{i=1}^{B} \to \boldsymbol{Z}=\{\boldsymbol{z}_i\}_{i=1}^{B}$. The supervised contrastive loss (SCL) function sums up the loss of each element of the mini-batch in $\boldsymbol{Z}$:
\begin{equation}
\mathcal{L}^{sup} = \sum_{i = 1}^{B} \mathcal{L}_{i}^{sup}.
\end{equation}

{In a mini-batch, suppose the positive output $\boldsymbol{z}_p \in \boldsymbol{Z}$ as the same category of $\boldsymbol{z}_i$ that does not contain $\boldsymbol{z}_i$, and the negative output $\boldsymbol{z}_a \in \boldsymbol{Z}$ as the arbitrary output except $\boldsymbol{z}_i$. The SCL function of each $z_i$ is:}
\begin{equation}   
\mathcal{L}_{i}^{sup} = \frac{-1}{|\boldsymbol{P}_i|} \sum_{p \in \boldsymbol{P}_i} \log \frac{\exp \left(\boldsymbol{z}_{i} \cdot \boldsymbol{z}_{p} / \tau\right)}{\sum\limits _{a \in \boldsymbol{A}_i} \exp \left(\boldsymbol{z}_{i} \cdot \boldsymbol{z}_{a} / \tau\right)},
\label{supcon loss}
\end{equation}
where $\boldsymbol{A}_i \equiv \{ \{1,2,\cdots, B\} \setminus  \left\{i\right\} \}$ is the indices of all mini-batch samples that do not contain $i$, $\boldsymbol{P}_i \equiv \left\{p \in \boldsymbol{A}_i:\tilde{{y}}_{p} = \tilde{{y}}_{i}\right\}$ is the indices of all mini-batch samples that have the same class as index $i$, $|\boldsymbol{P}_i|$ is the number of samples in $\boldsymbol{P}_i$, $\cdot$ denotes inner product, and $\tau \in \mathbb{R}^{+}$ is the temperature parameter. Therefore, $\frac{1}{|\boldsymbol{P}_i|}$ is used to calculate the average of the logarithmic term. Note that in supervised contrastive learning, the SCL function only uses the augmented samples to update network parameters. 


For any sample, all positives (augmented data with the same label) in a mini-batch contribute to the numerator. Thus, the goal of supervised contrastive learning is to minimize the SCL function, which increases the value of the numerator. This encourages the network to closely align representations to all instances from the same category, as updated parameters of the network make $\boldsymbol{z}_i \cdot \boldsymbol{z}_p$ become larger~\cite{khosla2020supervised, wang2021understanding}. 

As a type of representation learning, supervised contrastive learning also requires a classifier to achieve classification tasks. Two typical strategies include two-stage and one-stage training. Regarding the former, the first stage learns features by using the SCL function and the second stage updates classifiers using the cross-entropy loss function~\cite{khosla2020supervised}. As for the latter, the SCL function and the cross-entropy function are placed in two branches and trained simultaneously~\cite{wang2021contrastive}. The one-stage strategy showed efficiency and effectiveness in handling small data~\cite{yang2022few, peng2022progressively}, and we adopt this strategy for our framework.


However, supervised contrastive learning also faces difficulties in handling the imbalanced data \cite{zhang2022class}. In this case, the number of healthy samples is substantially larger than the faulty samples. In a mini-batch, healthy samples make up of the majority and other fault categories are only a minority or even non-existent. As a result, supervised contrastive learning struggles to increase the distance between minority classes, limiting its effectiveness in addressing the class imbalance. To overcome this issue, we consider enhancing the network's feature extraction capability and balancing the contrastive loss function within each class, thereby adapting it to handle long-tailed datasets.

\section{Proposed Methodology}
\begin{figure*}[pos=b]
\centering
\includegraphics[width=\textwidth]{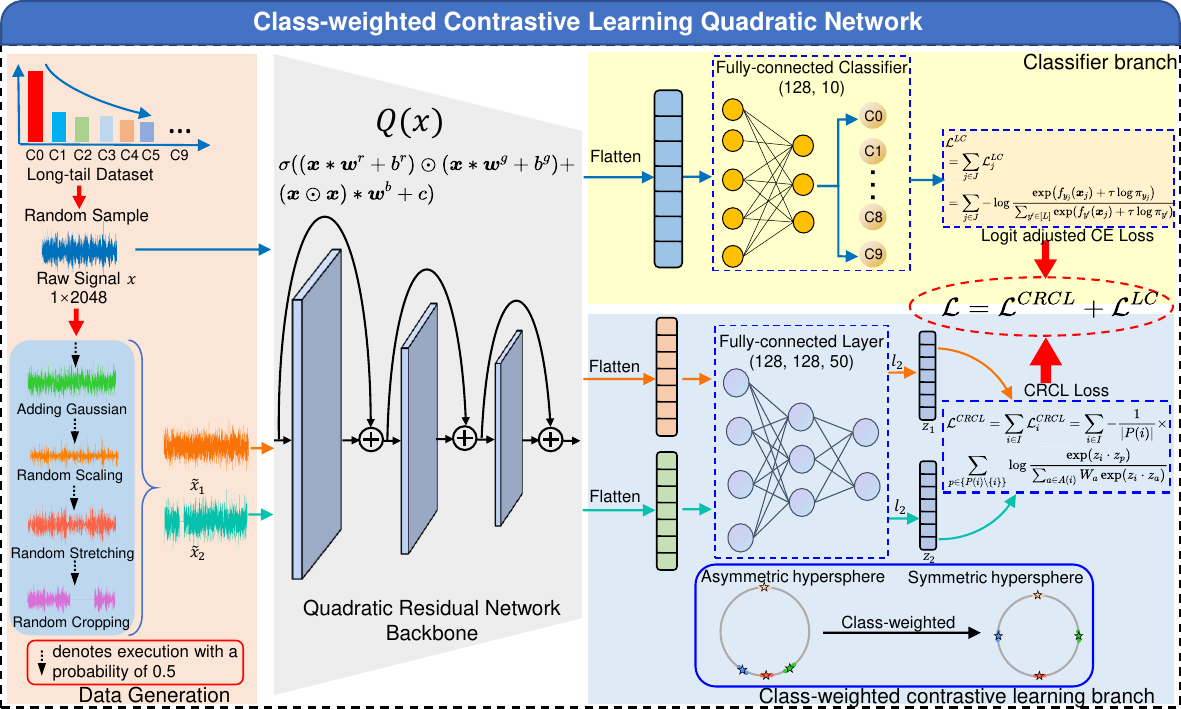}
\caption{An overview of the proposed framework}
\label{framework}
\end{figure*}

An overview of our proposed method is given in Figure~\ref{framework}, and the developed approach consists of four steps: data augmentation, quadratic network construction, classifier learning, and contrastive learning. The first step aims to generate extra data for contrastive learning by leveraging data augmentation techniques. The second step employs a quadratic convolutional residual network as the backbone of the model to extract informative features from the raw signal. In the third and fourth step, two MLPs are employed as projection networks: the classifier branch employs a logit-adjusted cross-entroy (CE) loss function to complete the classification task, while the contrastive learning branch utilizes a class-weighted loss function to capture the latent representation of long-tailed data.

\subsection{Data augmentation}
The proposed method starts by randomly sampling the long-tailed data, and  each selected raw sample $\boldsymbol{x}$ is passed through the data augmentation module to generate a pair of positive samples $\tilde{\boldsymbol{x}}_{1}$, $\tilde{\boldsymbol{x}}_{2}$ for supervised contrastive learning. There are four data augmentation methods commonly used in signal processing: 

\begin{enumerate}[listparindent=1.25em]
    \item \textbf{Adding Gaussian noise}: Gaussian noise $\boldsymbol{n}$ is added to each sample $\boldsymbol{x}$. Here, we set  $\boldsymbol{n}$ follows a Gaussian distribution $N(0, 0.01)$. Note that all data augmentation parameters we used refer to Ref. \cite{hu2021robust}.
    \begin{equation}
    \label{Add Gaussian Noise}
        \tilde{\boldsymbol{x}} = \boldsymbol{x} + \boldsymbol{n}. 
    \end{equation}

    \item \textbf{Random scaling}: A random variable $\boldsymbol{s}$ with value sampling from a Gaussian distribution $N(0, 0.01)$ is multiplied with each sample $\boldsymbol{x}$.
    \begin{equation}
        \label{Random Scale}
        \tilde{\boldsymbol{x}} = \boldsymbol{x} \cdot \boldsymbol{s}. 
    \end{equation}

    \item \textbf{Random stretching}: The sample $\boldsymbol{x}$ is stretched along the time axis. The degree of stretching is specified by a scaling factor denoted by $r$. The value of $r$ is randomly sampled from a uniform distribution $\mathrm{U}(0, 1)$ for each augmentation.  If the scaling factor is set to $r$, the resulting length of the stretched signal is $r$ times the length of the original signal $n$.

    Specifically, the stretched signal denoted $\boldsymbol{x}_\text{stretched}[j]$, is obtained by performing a linear interpolation of $\boldsymbol{x}$ at the positions $rj$, where $j$ is the index of the sample in the original signal. The final stretched signal is obtained by truncating the interpolated signal to the length of the original signal.
    \begin{equation}
        \label{Random Stretch}
        \begin{aligned}
        &\boldsymbol{x}_\text{stretched}[j]=\boldsymbol{x}[r j], \\
        &\tilde{\boldsymbol{x}}=\boldsymbol{x}_\text{stretched}[0:n].     
        \end{aligned}
    \end{equation}

    \item \textbf{Random cropping}: a random interval of the sample is set to zero. We set the interval length to 30 and replace it with a random position $j$ in a sample.
    \begin{equation}
        \label{Random Crop}
        \tilde{\boldsymbol{x}} = \{\boldsymbol{x}[0:j] , [\underbrace{0, 0, \cdots, 0}_{30}] , \boldsymbol{x}[j+30: \text{end}]\}.
    \end{equation}
\end{enumerate}

Two augmented sample pairs $\boldsymbol{\tilde{x}}_1, \boldsymbol{\tilde{x}}_2$ are produced by randomly selecting two augmentation methods, where each technique has an equal probability of 0.5 to be selected as the argumentation method. For example, $P(\tilde{\boldsymbol{x}}=\boldsymbol{x} + \boldsymbol{n})=0.5$. In doing this, we ensure the diversity in the generated contrastive samples, thus enhancing the model's generalizability.

\subsection{Quadratic residual network}
In this paper, we employ a quadratic convolutional network as the backbone of the neural network because a quadratic network exhibits a superior feature extraction ability compared with conventional neural networks \cite{fan2019quadratic, fan2021expressivity}. Specifically, the effectiveness of the quadratic network has been verified in bearing faults diagnosis \cite{liao2023attention}.

A quadratic network replaces conventional neurons with quadratic neurons composed of an inner product and a power term of the input vector. Mathematically, given an input sample {$\boldsymbol x \in \mathbb{R}^{1\times n}$},  $\boldsymbol x= \left[x_1, x_2, \cdots, x_n\right]$, a quadratic convolutional operation can be expressed as:
\begin{equation}
    Q(\boldsymbol x) = \sigma(\left(\boldsymbol{x}\ast \boldsymbol{w}^r+{b}^r \right) \odot \left( \boldsymbol{x}\ast \boldsymbol{w}^g+{b}^g \right) +\left( \boldsymbol{x}\odot \boldsymbol{x} \right) \ast \boldsymbol{w}^b+c),
\label{eq:qnn}
\end{equation}
where $\ast$ denotes the convolutional operation, $\odot$ denotes Hadamard product, $\boldsymbol w^r \in \mathbb{R}^{k\times 1}$, $\boldsymbol w^g \in \mathbb{R}^{k\times 1}$ and $\boldsymbol w^b \in \mathbb{R}^{k\times 1}$ denote the weights of three different convolution kernels, $\sigma(\cdot)$ is the activation function (e.g., ReLU), ${b}^r$, ${b}^g$ and ${c}$ denote biases corresponding to these convolution kernels.

{Mathematically, studies have shown that quadratic neurons exhibit a superior ability to approximate radial functions with polynomial-level neurons, whereas conventional neurons require exponential-level neurons \cite{fan2020universal, fan2021expressivity}. Additionally, quadratic networks can achieve polynomial approximation, but conventional neural networks can only achieve piece-wise approximation through non-linear activation functions. These characteristics have the potential to enhance the generalization and expressiveness of neural networks, as real-world data distributions are usually non-linear.}

{The advantage of quadratic neurons is that they can flexibly and directly enhance the performance of conventional networks, as networks are constructed by simply replacing conventional neurons. However, quadratic networks introduce more non-linear operations and parameters, leading to increased model complexity. It presents a challenge for achieving convergence during training.}


{To address the aforementioned issue, two strategies are implemented. First, quadratic networks are trained using an algorithm called ReLinear \cite{fan2021expressivity}. The parameters of quadratic neurons are factorized into two groups: the first-order group $\{\boldsymbol{w}^r, {b}^r\}$ and the quadratic group $\{\boldsymbol{w}^g, {b}^g, \boldsymbol{w}^b, c\}$. During the initialization stage, the first group undergoes normal initialization using Kaiming initialization \cite{he2015delving}, whereas the second group is set as $\{{b}^g=1; \boldsymbol{w}^g, \boldsymbol{w}^b, c = 0\}$. During the training stage, different learning rates, $\gamma_r$ and $\gamma_{g,b}$, are assigned to these two groups, with $\gamma_r = \alpha \gamma_{g,b}$, where $0<\alpha<1$. ReLinear initiates the training of quadratic network starting from the first-order terms and gradually trains the quadratic terms' parameters, enabling the neural network to prevent gradient explosion.}

\begin{figure}[pos=h]
\centering
\includegraphics[width=0.5\textwidth]{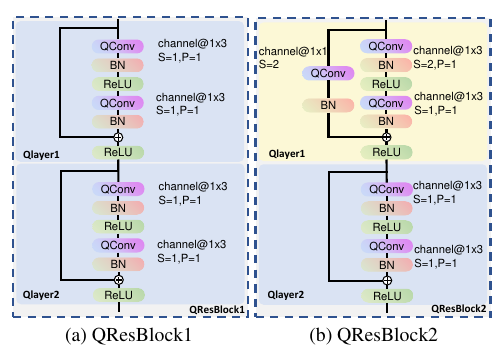}
\caption{The structure of quadratic ResNet backbone.}
\vspace{-0.5cm}
\label{QResNet}
\end{figure}

{Second, we construct a cross-layer connection strategy for the neural network to improve its stability. As depicted in Figure \ref{QResNet}, we employ two residual blocks, referred to as QResBlocks, each composed of two Qlayers. Each Qlayer contains the mainline $Q(\boldsymbol x)$ and a shortcut connection. To ensure compatibility with the channel dimension of the output variable, an additional "quadratic convolution-batch normalization" structure is introduced in Qlayer1 of QResBlock2, as illustrated in Figure \ref{QResNet} (b).
Overall, the structural parameters of the quadratic residual network backbone are presented in Table \ref{QResNet parameter}.}

\begin{table}[pos=h]
\centering
\caption{The structural parameters of quadratic ResNet backbone.}
\resizebox{\columnwidth}{!}{
\begin{tabular}{@{}ccccccc@{}}
\toprule
Number              & Type       & Kernel & Channel & Stride & Padding & Output  \\ \midrule
0                  & Input      & -      & -       & -      & -       & 1×2048  \\
\multirow{4}{*}{1} & QConv      & 1×7    & 16      & 2      & 3       & 16×1024 \\
                   & BN         & -      & -       & -      & -       & 16×1024 \\
                   & ReLU       & -      & -       & -      & -       & 16×1024 \\
                   & Max-Pool   & 1×3    & 16      & 2      & 1       & 16×512  \\
2                  & QResBlock1(2QConv) & 1×3      & 16      & 1      & 1       & 16×512  \\
3                  & QResBlock2(2QConv) & 1×3      & 32      & 1      & 1       & 32×256  \\
4                  & QResBlock2(2QConv) & 1×3      & 64      & 1      & 1       & 64×128  \\
5                  & QResBlock2(2QConv) & 1×3      & 128     & 1      & 1       & 128×64  \\
6                  & AvgPool    & -      & -       & -      & -       & 128×1   \\
7                  & Flatten    & -      & -       & -      & -       & 128     \\ \bottomrule
\end{tabular}}
\label{QResNet parameter}
\end{table}

\subsection{Class-weighted contrastive learning}
The two generated sample pairs are subsequently sent to the contrastive learning branch after passing through the backbone. In this branch, the features are transformed by an MLP and then mapped to a hypersphere using $L_2$ normalization (Eq. \eqref{eq:mapping}). The key idea of the class-weighted contrastive learning is to design the contrastive loss function to induce neural network to pay equal attention to each class.

In the case of a long-tailed dataset, the main impact of $\mathcal{L}^{sup}$ is attributed to the term $\sum\limits_ {a \in A(i)} \exp \left(\boldsymbol{z}_{i} \cdot \boldsymbol{z}_{a} / \tau\right)$, where $\boldsymbol{z}_{a}$ represents the output of each generated sample except for $\boldsymbol{z}_i$ \cite{zhu2022balanced}. When confronted with imbalanced datasets, tail classes pose a challenge for the network to effectively distinguish them from each other. This is because tail classes only occupy a small proportion of $\boldsymbol{z}_a$, even missing a certain tail category in some mini-batches. Consequently, the network's ability to accurately classify the tail classes is hindered by this imbalance in contribution between the head and tail classes.

Inspired by the reweighting technique, we design a class-aware weight to force the network to pay more attention to tail classes. That is,
\begin{equation}
    W_a = \frac{1}{|\boldsymbol{P}_a|},
\end{equation}
where $|\boldsymbol{P}_a|$ indicates the number of samples belonging to class $a$. A tail class $a$ has a much lower $|\boldsymbol{P}_a|$ and results in a larger $W_a$. {By doing this, each class has its own class weight, which makes the proportion of each class in the contrastive loss function balanced. As a result, the learned representation from contrastive learning remains unaffected by class imbalance.} 

Next, we integrate this weight into the SCL loss function formulated in Eq. (\eqref{supcon loss}), and we thus have the class-weighted contrastive loss (CRCL) as:
\begin{equation}
\begin{aligned}
&\mathcal{L}_{}^{CRCL} = \sum_{i = 1}^{B} \mathcal{L}_{i}^{CRCL} \\
= &\sum_{i = 1}^{B} -\frac{1}{\left|\boldsymbol{P}_i\right|} \sum_{p \in \boldsymbol{P}_i} \log \frac{\exp \left(\boldsymbol{z}_{i} \cdot \boldsymbol{z}_{p}\right)}{\sum_{a \in \boldsymbol{A}_i} W_a \exp \left(\boldsymbol{z}_{i} \cdot \boldsymbol{z}_{a}\right)}.
\label{bcl_loss}
\end{aligned}
\end{equation}

Compared to the original SCL loss function, for each $z_a$, we assign a weight $W_a$ for class $a$. Note that the temperature parameter $\tau$ is eliminated to reduce the number of hyperparameters that needs to be fine-tuned.

\begin{figure}[htbp]
    \centering
    \includegraphics{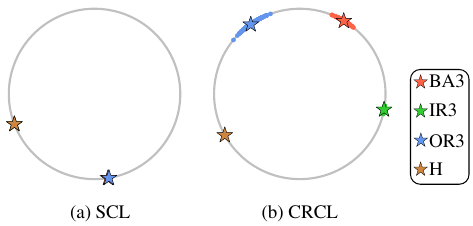}
    \caption{Features learned by SCL and CRCL on the CWRU dataset at IB rate=10:1. Where BA, IR, and OR denote ball defect, inner race defect, and outer race defect respectively.}
    \label{fig:cmp_scl_crcl}
\end{figure}

\textbf{Remark 1.} It has been proved that the SCL guides models collapsing to the vertices of a regular simplex lying on a hypersphere with balanced datasets, and empirical evidence indicates that a regular simplex configuration is beneficial to better performance \cite{graf2021dissecting}. We argue that class-weighted loss function encourages features corresponding to samples in the same class learned by the network has the equidistant from each other. This makes the feature center nearly located on the vertices of a regular simplex. {As shown in Figure \ref{fig:cmp_scl_crcl}, we map the imbalanced features acquired through SCL and CRCL onto a two-dimensional circle. It is evident that CRCL remarkably amplifies the separation between distinct classes. This attribute explains the ability of CRCL to handle imbalanced data.}

\subsection{Classifier learning}
Similar to contrastive learning, the raw signal is fed to the classifier branch after the backbone network. Here, we use the logit adjusted cross-entropy loss function $\mathcal{L}^{LC}$ to drive the training of the classifier \cite{menon2020long}. This function is also a class-weighted as shown below:
\begin{equation}
\mathcal{L}^{LC}= \sum_{j \in J}  \mathcal{L}^{LC}_j= \sum_{j \in J} -\log \frac{\exp \left(f_{y_j}(\boldsymbol x_j)+\tau \log \pi_{y_j}\right)}{\sum_{y^{\prime} \in[L]} \exp \left(f_{y^{\prime}}(\boldsymbol x_j)+\tau \log \pi_{y^{\prime}}\right)}.
\label{LC_loss}
\end{equation}
where $j \in J \equiv \{1, 2, \cdots, n\}$ are the indices of the raw data ${\boldsymbol x_1, \boldsymbol x_2, \cdots, \boldsymbol x_N}$, $f(\boldsymbol x_j)$ denotes the output of the classifier, i.e. logit. $f_{y_j}(\boldsymbol x_j)$ denotes the value of the element in the logit vector classified as label $y_j$. Let $[L]$ be a collection of labels $y'$ and $y' \in [L] \equiv \{1, 2, \cdots, L\}$, $\pi_{y_j}$ denotes the prior probability of the label $y_j$, and $\tau$ indicates temperature coefficient which is set to $\tau = 1$ in the developed approach.

Finally, we combine the $\mathcal{L}^{CRCL}$ and $\mathcal{L}^{LC}$ together to form the composite loss function $\mathcal{L}$:
\begin{equation}
    \mathcal{L} = \mathcal{L}^{CRCL} + \mathcal{L}^{LC}.
    \label{total_loss}
\end{equation}


\subsection{The superiority of quadratic networks}
Although some shreds of evidence suggest that the quadratic network is superior to its conventional counterpart in terms of feature extraction ability \cite{fan2023one}, while the key issue on \textit{why the quadratic network works better than conventional neural network when dealing with vibration signals} remains a mystery. In this subsection, we analyze the learning power of quadratic network and establish its connection with signal autocorrelation.

\subsubsection{Autocorrelation} 
Autocorrelation, also known as serial correlation in the discrete time case, measures the correlation of the signal using a delayed copy of itself. Essentially, it quantifies the similarity between two observations of a signal as a function of the time delay between them. It is used to find the periodic signal submerged by noise or identify the missing feature frequency in a signal\footnote{\url{https://en.wikipedia.org/wiki/Autocorrelation}}. Mathematically, the convolutional expression of autocorrelation $r(\boldsymbol{x}): \mathbb{R}^{1\times n} \to \mathbb{R}^{1 \times n}$ is as follows,
\begin{equation}
\begin{aligned}
     R(\boldsymbol{x}) &= \boldsymbol x \ast \boldsymbol x\\
     &= \left[ \sum_{i=1}^n{x_{i}^{2}}, \sum_{i=1}^{n-1}{x_ix_{i+1}},\cdots ,\sum_{i=1}^2{x_ix_{i+n-2}},x_1x_n \right]\\
     &=[r_1, r_2, \cdots, r_n].
\end{aligned}
\label{eq:acor}
\end{equation}

\textbf{Remark 2.} The convolution operation in signal processing contains both flip and translation operations, which is why autocorrelation has a negative sign ( $\boldsymbol{y}=-\boldsymbol{x}*\boldsymbol{x}$), but it only has the translation operation in deep learning. For the sake of formal uniformity, the convolution operation form of deep learning is used here. The advantage of autocorrelation is that it has excellent noise suppression. Suppose a noisy signal $\boldsymbol{y}$ consists of two parts:
\begin{equation}
    \boldsymbol{y} = \boldsymbol{x} + \boldsymbol{s},
\end{equation}
where $\boldsymbol x$ denotes the raw signal, $\boldsymbol s$ denotes the noise. 

An autocorrelation function calculates the noisy signal and its delay as:
\begin{equation}
\begin{aligned}
r_{yy, \tau} &= \sum_{i=1}^{n - \tau }{(x_i+s_i)(x_{i+\tau}+s_{i+\tau})} \\
&= \sum_{i=1}^{n - \tau }(x_ix_{i+\tau}+2s_ix_{i+\tau}+s_is_{i+\tau})\\
&= r_{xx,\tau}+2r_{xs,\tau}+r_{ss,\tau},
\end{aligned}
\end{equation}
where $r_{ab, \tau}$ denotes the value of the autocorrelation function of the signal $\boldsymbol{a}$ and $\boldsymbol{b}$ when the delay is $\tau$. As $\boldsymbol{s}$ is a random noise, it does not correlate with $\boldsymbol x$, so $r_{xs, \tau}=0$. If $\boldsymbol s$ itself is uncorrelated, then $r_{ss, \tau}=0$ except $\tau=0$. The following equation holds:
\begin{equation}
\left\{\begin{matrix}
r_{yy, \tau} =  r_{xx,\tau}+r_{ss,\tau}, \ \tau = 0\\
r_{yy, \tau} =  r_{xx,\tau}, \ \tau \ne 0.
\end{matrix}\right.
\end{equation}

The above equations indicate that the autocorrelation function can extract the feature of the signal from the noise.

\subsubsection{Learnable autocorrelation and quadratic neuron} 
We define learnable autocorrelation as multiplying a set of weight parameters $\mathcal{W}=\{\boldsymbol{W}_0,\boldsymbol{W}_1, \cdots, \boldsymbol{W}_n\}$ to autocorrelation. These parameters can be updated by backpropagation in a convolutional neural network: 
\begin{equation}
\begin{aligned}
&R^{\mathcal{W}}(\boldsymbol x) \\ &= \boldsymbol W  (\boldsymbol x \ast \boldsymbol x) \\
&= \left[\sum_{i=1}^n{w_{i}^1x_{i}^{2}}, \sum_{i=1}^{n-1}{w_{i}^2x_ix_{i+1}},\cdots ,\sum_{i=1}^2{w_{i}^{n-1}x_ix_{i+n-2}},w_1^nx_1x_n \right]\\
&=[r^{\boldsymbol{W}_0}_0,r^{\boldsymbol{W}_1}_1, \cdots,r^{\boldsymbol{W}_n}_n].
\end{aligned}
\label{eq:wac}
\end{equation}

When $\mathcal{W}\equiv \boldsymbol{1}$, the learnable autocorrelation degenerates to the conventional autocorrelation.


On the other hand, the output of a quadratic convolutional layer can be factorized as:
\begin{equation}
    Q(\boldsymbol x) = [q_1, q_2, \cdots, q_n],
\end{equation}
where
\begin{equation}
\begin{aligned}
q_j&=\left( \sum_{i=1}^k{w_{i}^{r}x_{i+j-1}+b^r} \right) \left( \sum_{i=1}^k{w_{i}^{g}x_{i+j-1}+b^g} \right) \\&+\sum_{i=1}^k{w_{i}^{b}x_{i+j-1}^{2}+c}
\\
=&\sum_{i=1}^k{\left( w_{i}^{r}w_{i}^{g}+w_{i}^{b} \right) x_{i+j-1}^{2}+}\sum_{i=1}^{k-1}{w_{i}^{r}w_{i+1}^{g}x_{i+j-1}x_{i+j}}
\\
+&\sum_{i=1}^{k-2}{w_{i}^{r}w_{i+2}^{g}x_{i+j-1}x_{i+j+1}}+\cdots +w_{1}^{r}w_{k}^{g}x_jx_{k+j-1}  
\\
+&b^r\sum_{i=1}^k{w_{i}^{g}x_{i+j}}+b^g\sum_{i=1}^k{w_{i}^{r}x_{i+j}}+C.
\end{aligned}
\label{eq:qexpand}
\end{equation}

It is obvious that the calculation of a quadratic convolutional neuron contains the learnable autocorrelation operation. Combining Eq. (\ref{eq:wac}) and Eq. 
 (\ref{eq:qexpand}), we have
\begin{equation}
\begin{aligned}
q_j=&\underset{{Learnable}\,\,{Autocorrelation}}{\underbrace{r_{0}^{\boldsymbol{W_0}}+r_{1}^{\boldsymbol{W_1}}+\cdots +r_{\boldsymbol{k}}^{\boldsymbol{W_k}}}}
\\+&\underset{{Covolutional}\,\,{Operation}}{\underbrace{b^r\sum_{i=1}^k{w_{i}^{g}x_{i+j}}+b^g\sum_{i=1}^k{w_{i}^{r}x_{i+j}}+C}}.
\end{aligned}
\label{eq:learnable_cor}
\end{equation}

\begin{figure}[pos=h]
\centering
\includegraphics[width=\linewidth]{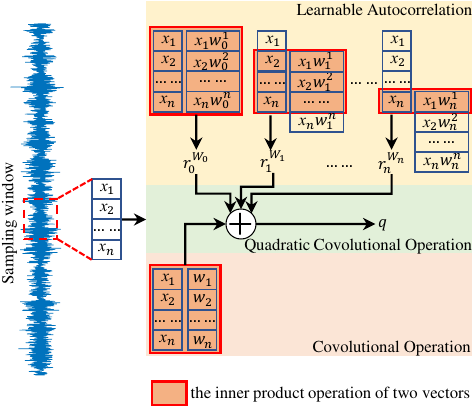}
\caption{The operation of the quadratic neuron.}
\label{q_autocorr}
\end{figure}

The quadratic convolutional operation can be decomposed into two parts: the sum of learnable autocorrelation and the conventional convolutional operation. As shown in Figure \ref{q_autocorr}, the first part is using the learnable autocorrelation for each subsequence adding to the final results. The second part is using learnable filters to convolute the input signal, which is the same as a conventional convolution neural network.

With the above deduction, a quadratic network offers advantages over conventional neural networks when processing signals. The autocorrelation operation within a quadratic neuron aids in extracting valuable signals with random noise, while such capability is missing in conventional neural networks. By leveraging this feature, our methodology is able to enhance the power of feature extraction from the input data, and thus leads to an improved performance in bearing fault diagnosis.


\section{Computational experiments}
In this section, we consider two bearing datasets: the Case
Western Reserve University (CWRU) rolling bearing dataset\footnote{\url{https://engineering.case.edu/bearingdatacenter/download-data-file}} and our own bearing dataset, to validate the performance of the proposed method. The two datasets consist of ten categories with nine faulty bearings and one healthy bearing. In the computational experiments, we first compare CCQNet with other state-of-the-art methods under a long-tailed distribution. Next, we analyze the feature extraction ability of the proposed model through visualization of interpretable feature maps and learnable autocorrelation. Finally, we conduct several ablation studies to verify the properties of our method.


\subsection{Dataset description}
\subsubsection{Case Western Reserve University bearing faults dataset (CWRU)} 
This extensively utilized dataset is collected by the Bearing Data Center at Case Western Reserve University (CWRU). The dataset is generated by applying electro-discharge machining (EDM) to motor deep groove ball bearings to artificially induce faults. Specifically, faults with diameters of 0.007 inches, 0.014 inches, and 0.021 inches are introduced into the inner race, outer race, and ball of the bearings, respectively. Subsequently, the faulty bearings are reinstalled in the test motors, and vibration data is recorded at both the drive-end (DE) and fan-end (FE). Throughout the experiments, the motor loads range from 0 to 3 HP (horsepower), while the motor speeds exhibit slight variations within the range of 1797 to 1720 RPM. The vibration signals are acquired at two sampling rates: 12 kHz and 48 kHz. In the experiments, we consider the vibration data collected at the drive-end (DE) with a 0 HP load and a sampling rate of 12 kHz.

\subsubsection{Our bearing dataset} 
\begin{figure}[pos=htbp]
    \centering
    \includegraphics[width=\linewidth]{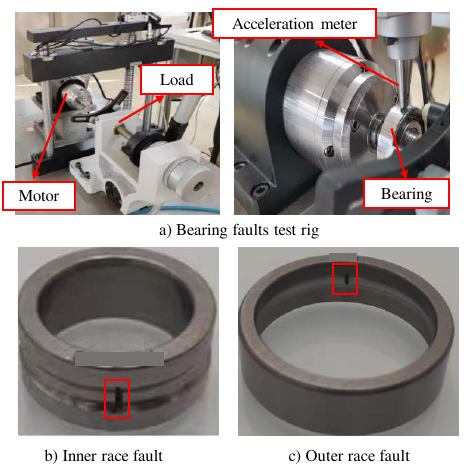}
    \caption{Our bearing faults test rig and bearing failure of inner race and outer race.}
    \label{fig:bearing}
\end{figure}

In addition to CWRU, we also collect our dataset employing angular contact ball bearings (HC7003), which are specifically designed for high-speed rotational machinery. This experiment is conducted at MIIT Key Laboratory of Aerospace Bearing Technology and Equipment, Harbin Institute of Technology. As depicted in Figure \ref{fig:bearing}, the acceleration meter is directly attached to the bearings to capture the vibration signals generated by the bearings. Consistent with the CWRU dataset, we induced faults at the outer race (OR), inner race (IR), and ball (BA), encompassing three levels of severity: slight, moderate, and severe. Table \ref{tab:ourdatasetfault} provides an overview of the fault types and the corresponding size in our dataset. During the test, the vibration signals were acquired at a constant motor speed of 1800 r/min using an NI USB-6002 device with a sampling rate of 12 kHz. We collected 47 seconds of bearing vibration data totaling 561,152 data points per fault category. Unlike the CWRU dataset, the bearing faults in our dataset consist of cracks with uniform sizes but varying depths. As a result, the vibration signals exhibit greater similarity among different fault types, thereby increasing the difficulty of accurately classifying the diagnostic model.

\begin{table}[pos=h]
\caption{Ten classes in our dataset. OR and IR denote that the faults appear in the outer race and inner race, respectively. Fault size denotes length×width×depth.}
\begin{tabular}{@{}llc@{}}
\toprule
Class & Fault mode              & Fault size (mm) \\ \midrule
C1    & Healthy Bearing         & 0×0×0                               \\
C2    & Ball cracking(Minor)    & 1×0.3×0.1                           \\
C3    & Ball cracking(Moderate) & 1×0.3×0.2                           \\
C4    & Ball cracking(Severe)   & 1×0.3×0.3                           \\
C5    & IR cracking(Minor)      & 2×0.3×0.1                           \\
C6    & IR cracking(Moderate)   & 2×0.3×0.2                           \\
C7    & IR cracking(Severe)     & 2×0.3×0.3                           \\
C8    & OR cracking(Minor)      & 2×0.3×0.1                           \\
C9    & OR cracking(Moderate)   & 2×0.3×0.2                           \\
C10   & OR cracking(Severe)     & 2×0.3×0.3                           \\ \bottomrule
\end{tabular}
\label{tab:ourdatasetfault}
\end{table}

\subsection{Experimental setup}
We conduct bearing faults diagnosis experiments by constructing long-tailed datasets. Here, we consider different imbalance rates (IB Rate) to investigate the variation of different models to the different levels of imbalance of the dataset. The IB Rate is defined as follows:
\begin{equation}
    \mathrm{IB\ Rate} = \frac{N_{normal}}{N_{fault}},
\end{equation}
where $N_{normal}$ and $N_{fault}$ represent the number of normal samples and the number of faulty samples in each class, respectively. 

{Next, due to each class of measured signals consisting of only one long sequence, we randomly extract a subsequence comprised of 2048 data points as one input $\boldsymbol{x}_i$. Afterwards, the entire dataset is divided into three parts: training set, validation set, and testing set. Specifically, in the training set, the number of normal data is set to 500, while the number of other fault categories is selected by the IB Rate (5:1, 10:1, 20:1, 50:1), which is set to 100, 50, 25, and 10, respectively. For all categories in the validation and test sets, we extracted 250 samples with balanced classes.} Identical standardization is applied to all the samples.

\subsubsection{Baseline methods}
We compare the developed method \footnote{Our code is available at \url{https://github.com/yuweien1120/CCQNet} for readers' verification.} with several well-known convolutional neural networks commonly used in the field of bearing fault diagnosis, including conventional CNN, contrastive learning-based methods, and resampling-based methods. The descriptions of baseline methods are given as follows:
\begin{enumerate}
    \item \textbf{WDCNN} is a widely used method for bearing fault diagnosis. It utilizes a wide convolutional kernel in the first convolutional layer to extract features and suppress noise \cite{zhang2017new}.
    \item \textbf{SupCon} is a supervised contrastive learning method. It leverages the concept of contrastive learning to enhance the discriminative power of the learned representations \cite{hu2021robust}.
    \item \textbf{SelfCon} is a self-supervised contrastive learning method. It applies contrastive learning in a self-supervised manner to learn effective representations \cite{chen2022self}. 
    \item \textbf{DNCNN} is a convolutional neural network that incorporates normalization techniques for both convolutional layers and neuron weights. This normalization helps improve the stability and generalization of the model \cite{jia2018deep}.
    \item \textbf{CA-SupCon} combines oversampling techniques (class-aware sampler) with supervised contrastive learning. This method aims to address class imbalance issues in the dataset and enhance the discriminative ability of the learned representations \cite{zhang2022class}.
    \item \textbf{Oversample+reweight} is a method that combines oversampling techniques with neuronal normalization for classifiers. It also adopts a reweighted loss function to tackle class imbalance problems in bearing fault diagnosis \cite{wu2021learning}.
\end{enumerate}

\subsubsection{Training strategies} 
We employ the stochastic gradient descent (SGD) optimizer to optimize all networks. The batch size is set to 64, and the training epoch is set to 200 for our method. Hyperparameters of other methods are set according to the values reported in the corresponding articles. In addition, we train CCQNet using the ReLinear algorithm \cite{fan2021expressivity}, which facilitates the fast convergence of quadratic networks. At the initial stage, the parameters in a quadratic neuron are initialized as $\boldsymbol{w}_g =0, \boldsymbol{w}_b = 0, c = 0$ and $b_g = 1$, while $\boldsymbol{w}_r$ and $b_r$ follow the random initialization. At the training stage, the linear terms and quadratic terms are trained with different learning rates with $\gamma_l$ and $\gamma_q$ respectively. We set a scale factor $\alpha$ to control the two learning rates, that is
\begin{equation}
    \gamma_q = \alpha \cdot \gamma_l, \ \ 0<\alpha<1.
\end{equation}

At last, the learning rate is adjusted using the cosine annealing strategy.

\subsubsection{Evaluation metrics} 
The classification performance of the model is assessed using accuracy (ACC), F1 score, and MCC (Matthews Correlation Coefficient). These metrics are chosen as they are insensitive to category imbalance. The definitions of these metrics are as follows:
\begin{equation}
    \mathrm{ACC} = \frac{1}{\mathrm{C}}  \sum^{\mathrm{C}}_{i=1} \frac{\mathrm{TP}_i}{\mathrm{TP}_i + \mathrm{FP}_i},
\label{acc}
\end{equation}

\begin{equation}
    \mathrm{F1} = \frac{1}{\mathrm{C}}  \sum^{\mathrm{C}}_{i=1} \frac{2\mathrm{TP}_i}{2\mathrm{TP}_i + \mathrm{FP}_i + \mathrm{FN}_i }, 
\label{f1}
\end{equation}

\begin{equation}
    \mathrm{MCC} = \frac{c \cdot s - \sum_{i=1}^{\mathrm{C}}p_i \cdot t_i}{\sqrt{\left (s^2 - \sum_{i=1}^{\mathrm{C}}p_i^2\right) \left (s^2 - \sum_{i=1}^{\mathrm{C}}t_i^2\right)} }.
\label{MCC}
\end{equation}
where $\mathrm{TP}_i$, $\mathrm{FP}_i$, and $\mathrm{FN}_i$ denote the true positive rate, false positive rate, and false negative rate for class $i$, respectively; $\mathrm{C}$ denotes the number of classes, $t_i$, $p_i$, $c$ and $s$ are the number of times class $i$ truly occurred, the number of times class $i$ is predicted, the total number of samples correctly predicted, and the total number of samples for $C$ classes respectively.





\subsection{Long-tail fault diagnosis results}
Tables \ref{performance cwru} and \ref{performance hit} report the performance of the considered models on CWRU dataset and our bearing dataset. Firstly, it is clear that when the IB rate rises, both ACC, F1, and MCC drop down. Even so, our method still outperforms other SOTA methods across all the metrics. Secondly, when the IB rate is small (5:1, 10:1), the advantage of our method is not obvious, i.e., WDCNN and CCQNet all reach 100\% in F1 and MCC. As the IB Rate rises, performance of all baseline methods drop significantly or even fail to classify. For instance, when IB rate = 50:1, oversample+reweight only has 19\% F1 on the CWRU dataset. However, CCQNet remains over 76\% F1 and MCC on two datasets at the 50:1 IB rate, while the second-best method WDCNN only achieves 67.18\% and 6.40\% in F1 on the two datasets, respectively. The results indicate that our proposed method excels at handling extremely imbalanced data.

\begin{table*}[pos=b]
\caption{Performance comparison (\%) on CWRU bearing dataset, where the bold-faced number indicates the best performance.}
\centering
\label{performance cwru}
\resizebox{2\columnwidth}{!}{
\begin{tabular}{@{}lcccccc@{}}
\toprule
IB rate             & \multicolumn{1}{c}{}     & \multicolumn{1}{c}{5:1}  & \multicolumn{1}{c}{}     & \multicolumn{1}{c}{}     & \multicolumn{1}{c}{10:1} & \multicolumn{1}{c}{}     \\ \midrule
                    & \multicolumn{1}{c}{ACC}  & \multicolumn{1}{c}{F1}   & \multicolumn{1}{c}{MCC}  & \multicolumn{1}{c}{ACC}  & \multicolumn{1}{c}{F1}   & \multicolumn{1}{c}{MCC}  \\ \midrule
WDCNN               & \textbf{100.00$\pm$0.00} & \textbf{100.00$\pm$0.00} & \textbf{100.00$\pm$0.00} & 99.85$\pm$0.06           & 99.85$\pm$0.06           & 99.83$\pm$0.07           \\
SupCon              & 99.61$\pm$0.09           & 95.09$\pm$0.09           & 99.57$\pm$0.10           & 93.36$\pm$0.34           & 92.39$\pm$0.47           & 92.95$\pm$0.34           \\
SelfCon             & 95.24$\pm$0.40           & 95.09$\pm$0.44           & 94.82$\pm$0.43           & 86.30$\pm$0.54           & 84.64$\pm$0.66           & 85.51$\pm$0.56           \\
DNCNN               & 94.43$\pm$0.52           & 94.26$\pm$0.56           & 94.03$\pm$0.54           & 81.95$\pm$0.73           & 80.59$\pm$0.91           & 80.48$\pm$0.76           \\
CA-Supcon           & 99.82$\pm$0.08           & 99.82$\pm$0.08           & 99.80$\pm$0.09           & 96.42$\pm$0.30           & 96.37$\pm$0.31           & 96.09$\pm$0.32           \\
oversample+reweight & 91.67$\pm$0.13           & 89.80$\pm$0.22           & 91.41$\pm$0.12           & 79.96$\pm$0.14           & 73.86$\pm$0.22           & 79.25$\pm$0.14           \\
CCQNet              & \textbf{100.00$\pm$0.00} & \textbf{100.00$\pm$0.00} & \textbf{100.00$\pm$0.00} & \textbf{100.00$\pm$0.00} & \textbf{100.00$\pm$0.00} & \textbf{100.00$\pm$0.00} \\ \midrule
IB rate             & \multicolumn{1}{c}{}     & \multicolumn{1}{c}{20:1} & \multicolumn{1}{c}{}     & \multicolumn{1}{c}{}     & \multicolumn{1}{c}{50:1} & \multicolumn{1}{c}{}     \\ \midrule
WDCNN               & 93.66$\pm$0.46           & 93.26$\pm$0.52           & 93.14$\pm$0.49           & 69.3$\pm$0.52            & 67.18$\pm$0.65           & 66.81$\pm$0.55           \\
SupCon              & 78.72$\pm$0.25           & 72.88$\pm$0.19           & 78.03$\pm$0.30           & 59.92$\pm$0.21           & 50.74$\pm$0.30           & 59.78$\pm$0.25           \\
SelfCon             & 75.23$\pm$0.26           & 73.23$\pm$0.26           & 73.33$\pm$0.27           & 41.63$\pm$0.32           & 36.55$\pm$0.31           & 37.13$\pm$0.37           \\
DNCNN               & 71.19$\pm$0.41           & 70.85$\pm$0.39           & 68.35$\pm$0.47           & 61.24$\pm$0.83           & 59.61$\pm$0.94           & 57.79$\pm$0.90           \\
CA-Supcon           & 82.21$\pm$0.39           & 81.77$\pm$0.42           & 80.56$\pm$0.42           & 62.65$\pm$0.52           & 58.95$\pm$0.54           & 59.66$\pm$0.57           \\
oversample+reweight & 65.37$\pm$0.44           & 60.37$\pm$0.44           & 64.07$\pm$0.49           & 30.05$\pm$0.06           & 19.00$\pm$0.11           & 25.07$\pm$0.10           \\
CCQNet              & \textbf{99.37$\pm$0.12}  & \textbf{99.37$\pm$0.12}  & \textbf{99.31$\pm$0.12}  & \textbf{80.12$\pm$0.34}  & \textbf{76.59$\pm$0.38}  & \textbf{78.93$\pm$0.38}  \\ \bottomrule
\end{tabular}}
\end{table*}

\begin{table*}[pos=t]
\caption{Performance comparison (\%) on our bearing dataset, where the bold-faced number indicates the best performance.}
\centering
\label{performance hit}
\resizebox{2\columnwidth}{!}{
\begin{tabular}{@{}lcccccc@{}}
\toprule
IB rate             & \multicolumn{1}{c}{}    & \multicolumn{1}{c}{5:1}  & \multicolumn{1}{c}{}    & \multicolumn{1}{c}{}    & \multicolumn{1}{c}{10:1} & \multicolumn{1}{c}{}    \\ \midrule
                    & \multicolumn{1}{c}{ACC} & \multicolumn{1}{c}{F1}   & \multicolumn{1}{c}{MCC} & \multicolumn{1}{c}{ACC} & \multicolumn{1}{c}{F1}   & \multicolumn{1}{c}{MCC} \\ \midrule
WDCNN               & 87.74$\pm$0.34          & 87.46$\pm$0.31           & 86.58$\pm$0.37          & 66.31$\pm$3.39          & 65.02$\pm$3.32           & 63.23$\pm$3.63          \\
SupCon              & 93.36$\pm$0.35          & 92.84$\pm$0.40           & 92.83$\pm$0.39          & 83.05$\pm$3.45          & 82.00$\pm$3.71           & 81.69$\pm$3.69          \\
SelfCon             & 93.84$\pm$0.32          & 93.66$\pm$0.34           & 93.25$\pm$0.34          & 57.81$\pm$1.59          & 50.74$\pm$2.33           & 55.55$\pm$1.62          \\
DNCNN               & 85.80$\pm$0.56          & 85.34$\pm$0.58           & 84.37$\pm$0.61          & 78.18$\pm$0.48          & 77.13$\pm$0.58           & 76.05$\pm$0.52          \\
CA-Supcon           & 93.04$\pm$0.32          & 92.67$\pm$0.37           & 92.44$\pm$0.35          & 88.81$\pm$0.73          & 88.37$\pm$0.78           & 87.73$\pm$0.79          \\
oversample+reweight & 86.47$\pm$0.43          & 84.75$\pm$0.48           & 85.73$\pm$0.45          & 88.36$\pm$1.29          & 87.26$\pm$1.44           & 87.38$\pm$1.40          \\
CCQNet              & \textbf{96.69$\pm$0.19} & \textbf{96.64$\pm$0.20}  & \textbf{96.37$\pm$0.21} & \textbf{93.27$\pm$0.34} & \textbf{93.03$\pm$0.35}  & \textbf{92.73$\pm$0.37} \\ \midrule
IB rate             & \multicolumn{1}{c}{}    & \multicolumn{1}{c}{20:1} & \multicolumn{1}{c}{}    & \multicolumn{1}{c}{}    & \multicolumn{1}{c}{50:1} & \multicolumn{1}{c}{}    \\ \midrule
WDCNN               & 40.96$\pm$1.98          & 38.04$\pm$2.31           & 36.15$\pm$2.25          & 12.65$\pm$0.24          & 6.40$\pm$0.37            & 4.99$\pm$0.40           \\
SupCon              & 38.83$\pm$2.62          & 32.21$\pm$2.64           & 34.64$\pm$2.83          & 16.88$\pm$0.47          & 10.36$\pm$0.31           & 10.45$\pm$0.69          \\
SelfCon             & 34.94$\pm$3.10          & 25.65$\pm$2.58           & 31.27$\pm$3.87          & 11.00$\pm$0.85          & 6.88$\pm$0.53            & 1.24$\pm$1.06           \\
DNCNN               & 41.45$\pm$0.82          & 39.65$\pm$0.76           & 35.59$\pm$0.90          & 14.99$\pm$0.87          & 10.76$\pm$1.16           & 7.25$\pm$1.16           \\
CA-Supcon           & 50.33$\pm$2.58          & 47.12$\pm$2.99           & 45.99$\pm$2.80          & 16.04$\pm$1.43          & 12.66$\pm$2.01           & 7.76$\pm$1.70           \\
oversample+reweight & 59.50$\pm$1.06          & 54.39$\pm$1.66           & 56.94$\pm$1.14          & 10.32$\pm$0.30          & 3.40$\pm$0.37            & 0.41$\pm$0.39           \\
CCQNet              & \textbf{92.39$\pm$0.34} & \textbf{91.16$\pm$0.41}  & \textbf{91.83$\pm$0.33} & \textbf{75.25$\pm$1.11} & \textbf{72.64$\pm$1.16}  & \textbf{73.31$\pm$1.22} \\ \bottomrule
\end{tabular}}
\end{table*}

\begin{figure*}[h]
\centering
\includegraphics[width=\textwidth]{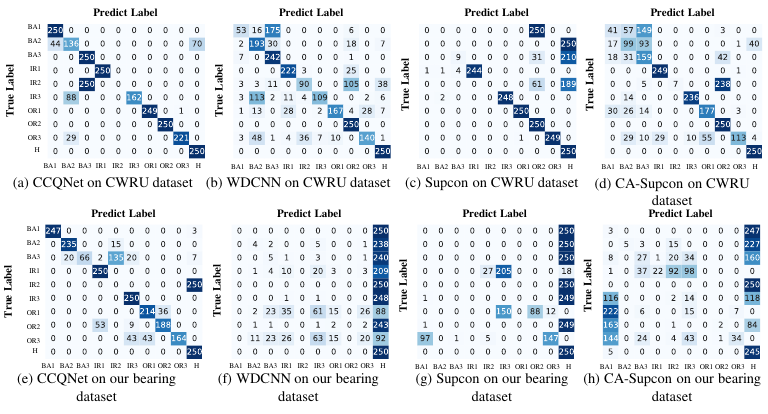}
\caption{The confusion matrices of different methods at IB rate = 50:1.}
\label{cm}
\vspace{-0.5cm}
\end{figure*}

Next, we focus on evaluating the performance of the top 4 best-performing methods at an imbalance rate of 50:1. To this end, we conduct a detailed analysis of their classification capability for each fault mode by examining the confusion matrix. As shown in Figure \ref{cm}, all the methods possess basic fault detection capabilities while their ability to accurately classify the different fault modes is affected by the imbalance level. It is noteworthy that all methods achieve high accuracy when classifying the healthy category, which can be attributed to the sufficient number of healthy samples available to train the model. However, in the CWRU dataset and our own bearing dataset, the three baseline methods exhibit poor performance in classifying ball faults (BA1, BA2, BA3) and inner race faults (IR2, IR3), respectively. In contrast, our proposed CCQNet demonstrates a much better classification accuracy for these fault modes. Additionally, CCQNet exhibits superior performance in the overall fault classification performance compared to the other three methods. These observations suggest that CCQNet is better equipped to handle extreme class imbalance conditions. 

\subsection{Classification visualization}
\subsubsection{Visualization by t-SNE}
\begin{figure*}[t]
\centering
\includegraphics[width=\textwidth]{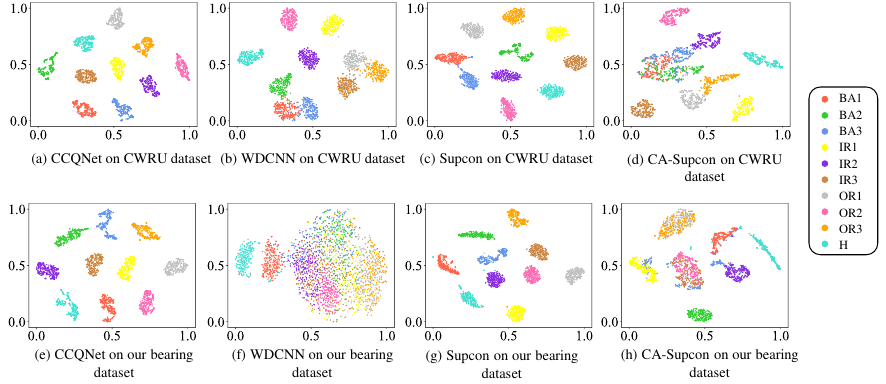}
\caption{t-SNE visualization of different methods at 50:1 IB Rate, where BA denotes ball defect, IR denotes inner race defect, OR denotes outer race defect, and H denotes health.}
\vspace{-0.5cm}
\label{tsne}
\end{figure*}

We use t-distributed stochastic neighbor embedding (t-SNE) to visualize the learned feature representation in a 2D space. For the sake of comparison, we only select the first four best methods for comparison. Figure \ref{tsne} shows 2D feature maps on each test set at 100:1 IB Rate. Firstly, regarding the CWRU dataset, the ball faults (BA1, BA2, BA3) of the three compared methods are separated insufficiently, which can lead to misdiagnosing. Secondly, in our dataset, the same phenomenon appears in some clusters, the clusters of the three compared methods all show an overlap. However, it can be seen that the clusters of different classes are more evenly distributed in CCQNet, and it is possible to better separate the inter-class samples and better aggregate the intra-class samples. The results also suggest that CCQNet has better feature extraction ability under extreme class imbalance conditions. 

\subsubsection{Feature map visualization}
\begin{figure}[t]
\centering
\includegraphics[width=\columnwidth]{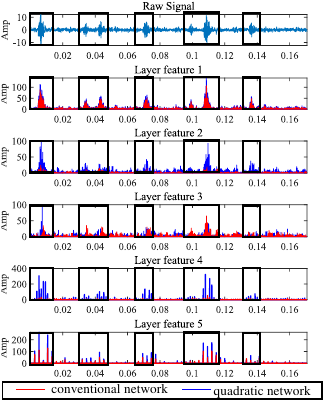}
\caption{Comparison of the features of each layer of the first-order and quadratic network.}
\vspace{-0.5cm}
\label{feature}
\end{figure}

Here, we provide insights into the superior feature extraction capabilities of quadratic networks compared to their conventional counterparts. Figure \ref{feature} show cases the feature maps at each layer of our quadratic residual network and a corresponding ResNet backbone with the same structure. Firstly, it is evident that the input signal exhibits distinct local features of high amplitude, resulting from fault-induced vibrations. Notably, both the conventional and quadratic networks preserve these local features at each layer. However, as the network layers become deeper, the quadratic network progressively places more emphasis on the local features with high amplitude values in the raw signal, surpassing the conventional network. Particularly, from the second to the fifth layer of visualized features, the quadratic network gives heightened attention to local features associated with fault-related signals. This finding strongly indicates that quadratic networks have superior feature representation to support accurate fault diagnosis.

\subsubsection{Learnable autocorrelation}
\begin{figure}[t]
\includegraphics[width=\columnwidth]{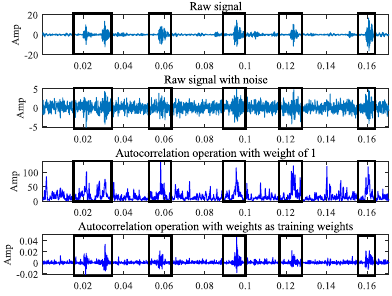}
\caption{Comparison of weighted autocorrelation operations with a weight of 1 and learnable autocorrelation operations with training weights in the time and time-frequency domains.}
\label{autocorr compare}
\end{figure}

Consistent with our earlier analysis, the effectiveness of quadratic networks lies in their ability to perform learnable autocorrelation, and enables extraction of significant signal features from noise. To investigate the practical anti-noise robustness of quadratic networks, we conduct experiments by introducing Gaussian noise with a variance of 10 to the signals. Specifically, we visualized the autocorrelation operation term that weights set as 1 and the learnable autocorrelation term (Eq. \eqref{eq:learnable_cor}) in the first layer of a quadratic network. The results as shown in Figure \ref{autocorr compare} indicate that while the autocorrelation operation can effectively extract fault-related signals from the noise, it also amplifies certain parts of extraneous noise. This limitation arises because autocorrelation is primarily designed to enhance transient impulses in the signal. In contrast, the learnable autocorrelation term demonstrates a remarkable ability to extract fault-related signals. The weights learned by the neural network act as adaptive filters, which further enhances the feature extraction capability. These findings emphasize the crucial role of learnable autocorrelation terms in empowering quadratic networks for effective signal feature extraction.

\subsection{Analysis experiments}


\subsubsection{Hyperparameter sensitivity}
It is crucial to consider the hyperparameters of the quadratic network, particularly the scale factor $\alpha$. In this study, we investigated the effects of $\alpha$ on the model's performance and conducted experiments to evaluate its sensitivity. The results are reported in Table \ref{Hyperparameter Sensitivity}. When the IB rate is set to 50:1, we observed that using an $\alpha$ value below $10^{-3}$ led to improved results. However, when $\alpha$ exceeded $10^{-3}$, the model parameters were unable to converge fully, resulting in a decrease in performance. Therefore, for our dataset, the optimal $\alpha$ value is determined to be 0.00001, which achieved an accuracy (ACC) of 75.25\%. For the CWRU dataset, the optimal $\alpha$ value is set to be 0.00005, resulting in an ACC of 80.12\%. Note that the exact value of $\alpha$ may vary depending on the datasets. Therefore, we recommend conducting thorough experiments to determine the optimal $\alpha$.

\begin{table}[pos=h]
\normalfont
\caption{The sensitivity of $\alpha$ to accuracy (\%) on the CWRU and our bearing dataset at 50:1 IB rate.}
\label{Hyperparameter Sensitivity}
\resizebox{1\columnwidth}{!}{
\begin{tabular}{cl|cccccc}
\toprule
\multicolumn{2}{c|}{$\alpha$}  & 0.005  & 0.001  & 0.0005 & 0.0001         & 0.00005 & 0.00001         \\ \midrule
\multicolumn{2}{c|}{Our dataset}  & 21.86 & 44.21 & 49.89  & 66.78 & \textbf{75.25}   & 64.34          \\
\multicolumn{2}{c|}{CWRU} & 66.94 & 51.06  & 68.72 & 60.78          & 55.49  & \textbf{80.12} \\ \bottomrule
\end{tabular}}
\end{table}

\subsubsection{Discussions of quadratic networks}
\textbf{Quadratic neurons vs conventional neurons.} Here we conduct experiments to verify the efficiency of quadratic networks. We use quadratic ResNet (QResNet), and two conventional ResNet models (one with the same structure as QResNet and ResNet34) as the backbones. The experiments were conducted with an imbalance rate of 100:1, and the results are presented in Table \ref{different backbones}. Despite ResNet34 having a larger number of parameters, the superior performance of QResNet demonstrates the efficiency and effectiveness of quadratic networks in bearing fault diagnosis. It is evident that QResNet achieves an accuracy of 75.25\% on our dataset and 80.12\% on the CWRU dataset, compared to 71.15\% and 69.33\% achieved by ResNet34, respectively. Additionally, the ResNet model with the same structure as QResNet exhibits the worst performance, with an accuracy that is 22.94\% lower on our dataset and 13.83\% lower on the CWRU dataset compared to QResNet. This further highlights the advantages of using a quadratic network in achieving better classification performance and higher computational efficiency.

\begin{table}[pos=h]
\normalfont
\caption{The accuracy (\%) of quadratic ResNet and conventional ResNet backbones on the CWRU and our dataset at 50:1 IB rate.}
\label{different backbones}
\resizebox{1\columnwidth}{!}{
\begin{tabular}{l|cccc}
\toprule
Method   & \multicolumn{1}{l}{\#Params} & \multicolumn{1}{l}{\#Flops} & \multicolumn{1}{l}{Our bearing dataset ACC(\%)} & \multicolumn{1}{l}{CWRU ACC(\%)} \\ \midrule
ResNet   & 200K                        & 22.6M                      & 52.31                          & 66.29                            \\
ResNet34 & 7200K                       & 719M                       & 71.15                          & 69.33                            \\
QResNet  & \textbf{700K}               & \textbf{66M}               & \textbf{75.25}                  & \textbf{80.12}                   \\ \bottomrule
\end{tabular}}
\end{table}

\textbf{The form of quadratic neurons.} To investigate the contributions of different parts of quadratic neurons, we conduct experiments using two degraded versions of the standard quadratic neuron. As shown in Table \ref{tab:qneurons}, we construct two different types of quadratic neurons, which are degraded versions of the standard quadratic neuron. In the first one, we remove the power term, and in the second one, we remove one of its inner-product terms.  As shown in Table \ref{tab:qneurons}, the computational results clearly suggest that the standard quadratic neuron outperforms the degraded versions. In particular, the accuracy of the standard one is roughly 10\% higher than that of its degraded counterparts on our dataset. Secondly, the performance of two degraded neurons exhibits variations across the two datasets. The degraded neuron lacking an inner-product term shows a slight decrease in accuracy (79.39\%) on the CWRU dataset but experiences a significant drop on our dataset (62.16\%). On the other hand, the degraded neuron without the power term demonstrates a remarkable decrease in performance on both datasets, and achieves an accuracy of 75.30\% and 62.38\% on the CWRU and our dataset, respectively. The results indicate that the power term of a quadratic neuron plays an important role in the network.  
\begin{table}[pos=h]
\normalfont
\caption{The accuracy (\%) of different quadratic functions on the CWRU and our dataset at 50:1 IB rate.}
\label{different quadratic functions}
\resizebox{1\columnwidth}{!}{
\begin{tabular}{cl|cc}
\toprule
\multicolumn{2}{c|}{Quadratic Function}                                                                                                                                                                                                    & CWRU    & Our dataset    \\ \midrule
\multicolumn{2}{c|}{$\left(\boldsymbol{x}\ast \boldsymbol{w}^r+{b}^r \right) \odot \left( \boldsymbol{x}\ast \boldsymbol{w}^g+{b}^g \right)$}                                                                            & 75.30          & 62.38         \\
\multicolumn{2}{c|}{$\left(\boldsymbol{x}\ast \boldsymbol{w}^r+{b}^r \right) + \left( \boldsymbol{x}\odot \boldsymbol{x} \right) \ast \boldsymbol{w}^b+c$}                                                               & 79.39          & 62.16         \\
\multicolumn{2}{c|}{$\left(\boldsymbol{x}\ast \boldsymbol{w}^r+{b}^r \right) \odot \left( \boldsymbol{x}\ast \boldsymbol{w}^g+{b}^g \right) +\left( \boldsymbol{x}\odot \boldsymbol{x} \right) \ast \boldsymbol{w}^b+c$} & \textbf{80.12} & \textbf{75.25} \\ \bottomrule
\end{tabular}}
\label{tab:qneurons}
\end{table}

\textbf{Quadratic network structures.} We perform experiments at IB Rate = 50:1 to assess the effect of the number of layers and the kernel size of the quadratic network on the model performance. The results are summarized in Tables \ref{number of model layers} and \ref{different kernel size}. With regard to the former, it can be observed that increasing the number of layers in the quadratic network leads to an improved performance. When the number of model layers is 5, the network achieves an accuracy of 75.25\% on our dataset and an accuracy of 80.12\% on the CWRU dataset.  As for the latter, the optimal kernel size is 7, which yields the best performance on both datasets. Based on these results, it can be concluded that a quadratic network with five layers and a convolutional kernel size of 7 is the optimal structure for fault diagnosis. 

\begin{table}[pos=h]
\normalfont
\caption{Accuracy (\%) of using different model layers with CCQNet on the CWRU and our dataset at 50:1 IB rate.}
\label{number of model layers}
\resizebox{1\columnwidth}{!}{
\begin{tabular}{c|cccc}
\toprule
Number of model layers & 2     & 3     & 4     & 5              \\ \midrule
CWRU                   & 74.16 & 78.1 & 73.59 & \textbf{80.12} \\
Our dataset                    & 51.77 & 44.4 & 53.2 & \textbf{75.25} \\ \bottomrule
\end{tabular}}
\end{table}

\begin{table}[pos=h]
\normalfont
\caption{Accuracy (\%) of using different kernel sizes with CCQNet on the CWRU and our dataset at 50:1 IB rate.}
\label{different kernel size}
\resizebox{1\columnwidth}{!}{
\begin{tabular}{c|ccccc}
\toprule
kernel size & 3     & 7              & 9    & 13    & 21    \\ \midrule
CWRU        & 78.76 & \textbf{80.12} & 78.49 & 77.44 & 71.1 \\
Our dataset         & 64.76 & \textbf{75.25} & 47.81 & 57.83 & 56.62 \\ \bottomrule
\end{tabular}}
\end{table}

\section{Conclusion}
In this paper, we have introduced a novel framework called CCQNet, based on quadratic networks, for addressing long-tailed distribution scenarios in bearing fault diagnosis. CCQNet incorporates a quadratic residual network (QResNet) as its backbone for effective feature extraction and employs two branches with class-weighted contrastive loss and logit-adjusted cross-entropy loss to ensure equal attention to all categories. We have also provided insights into quadratic networks, highlighting their ability to focus on local fault signal features through autocorrelation operations. Experimental results have demonstrated the outstanding performance of CCQNet in highly imbalanced data scenarios. Our work promotes the application of contrastive learning for long-tailed distribution-based bearing fault diagnosis and encourages further exploration of quadratic neural networks. As CCQNet serves as a versatile framework for long-tailed data classification, future research should explore its potential in a broader range of application scenarios. {For instance, extending it to few-shot or zero-shot learning, which are also highly desirable issues in fault diagnosis.}

\section*{Acknowledgement}
We would like to express our sincere gratitude to Prof. Xiaoli Zhao and Mr Hongyuan Zhang from the MIIT Key Laboratory of Aerospace Bearing Technology and Equipment, Harbin Institute of Technology for their selfless support and valuable assistance in conducting the bearing experiments for this research. Their contributions have been invaluable and greatly appreciated.

\bibliographystyle{model1-num-names}

\bibliography{reference}

\begin{thebibliography}{54}
\expandafter\ifx\csname natexlab\endcsname\relax\def\natexlab#1{#1}\fi
\providecommand{\url}[1]{\texttt{#1}}
\providecommand{\href}[2]{#2}
\providecommand{\path}[1]{#1}
\providecommand{\DOIprefix}{doi:}
\providecommand{\ArXivprefix}{arXiv:}
\providecommand{\URLprefix}{URL: }
\providecommand{\Pubmedprefix}{pmid:}
\providecommand{\doi}[1]{\href{http://dx.doi.org/#1}{\path{#1}}}
\providecommand{\Pubmed}[1]{\href{pmid:#1}{\path{#1}}}
\providecommand{\bibinfo}[2]{#2}
\ifx\xfnm\relax \def\xfnm[#1]{\unskip,\space#1}\fi
\bibitem[{Wang et~al.(2020)Wang, Shen, Xia, Wang, Zhu, and Zhu}]{wang2020multi}
\bibinfo{author}{X.~Wang}, \bibinfo{author}{C.~Shen}, \bibinfo{author}{M.~Xia},
  \bibinfo{author}{D.~Wang}, \bibinfo{author}{J.~Zhu},
  \bibinfo{author}{Z.~Zhu},
\newblock \bibinfo{title}{Multi-scale deep intra-class transfer learning for
  bearing fault diagnosis},
\newblock \bibinfo{journal}{Reliability Engineering \& System Safety}
  \bibinfo{volume}{202} (\bibinfo{year}{2020}) \bibinfo{pages}{107050}.
\bibitem[{Cao et~al.(2021)Cao, Ding, Jia, and Tian}]{cao2021novel}
\bibinfo{author}{Y.~Cao}, \bibinfo{author}{Y.~Ding}, \bibinfo{author}{M.~Jia},
  \bibinfo{author}{R.~Tian},
\newblock \bibinfo{title}{A novel temporal convolutional network with residual
  self-attention mechanism for remaining useful life prediction of rolling
  bearings},
\newblock \bibinfo{journal}{Reliability Engineering \& System Safety}
  \bibinfo{volume}{215} (\bibinfo{year}{2021}) \bibinfo{pages}{107813}.
\bibitem[{Vencl and Rac(2014)}]{vencl2014diesel}
\bibinfo{author}{A.~Vencl}, \bibinfo{author}{A.~Rac},
\newblock \bibinfo{title}{Diesel engine crankshaft journal bearings failures:
  case study},
\newblock \bibinfo{journal}{Engineering failure analysis} \bibinfo{volume}{44}
  (\bibinfo{year}{2014}) \bibinfo{pages}{217--228}.
\bibitem[{Islam and Kim(2019)}]{islam2019reliable}
\bibinfo{author}{M.~M. Islam}, \bibinfo{author}{J.-M. Kim},
\newblock \bibinfo{title}{Reliable multiple combined fault diagnosis of
  bearings using heterogeneous feature models and multiclass support vector
  machines},
\newblock \bibinfo{journal}{Reliability Engineering \& System Safety}
  \bibinfo{volume}{184} (\bibinfo{year}{2019}) \bibinfo{pages}{55--66}.
\bibitem[{Ali et~al.(2015)Ali, Chebel-Morello, Saidi, Malinowski, and
  Fnaiech}]{ali2015accurate}
\bibinfo{author}{J.~B. Ali}, \bibinfo{author}{B.~Chebel-Morello},
  \bibinfo{author}{L.~Saidi}, \bibinfo{author}{S.~Malinowski},
  \bibinfo{author}{F.~Fnaiech},
\newblock \bibinfo{title}{Accurate bearing remaining useful life prediction
  based on weibull distribution and artificial neural network},
\newblock \bibinfo{journal}{Mechanical Systems and Signal Processing}
  \bibinfo{volume}{56} (\bibinfo{year}{2015}) \bibinfo{pages}{150--172}.
\bibitem[{Smith and Randall(2015)}]{smith2015rolling}
\bibinfo{author}{W.~A. Smith}, \bibinfo{author}{R.~B. Randall},
\newblock \bibinfo{title}{Rolling element bearing diagnostics using the case
  western reserve university data: A benchmark study},
\newblock \bibinfo{journal}{Mechanical Systems and Signal Processing}
  \bibinfo{volume}{64} (\bibinfo{year}{2015}) \bibinfo{pages}{100--131}.
\bibitem[{Randall and Antoni(2011)}]{randall2011rolling}
\bibinfo{author}{R.~B. Randall}, \bibinfo{author}{J.~Antoni},
\newblock \bibinfo{title}{Rolling element bearing diagnostics—a tutorial},
\newblock \bibinfo{journal}{Mechanical Systems and Signal Processing}
  \bibinfo{volume}{25} (\bibinfo{year}{2011}) \bibinfo{pages}{485--520}.
\bibitem[{Randall et~al.(2001)Randall, Antoni, and
  Chobsaard}]{randall2001relationship}
\bibinfo{author}{R.~B. Randall}, \bibinfo{author}{J.~Antoni},
  \bibinfo{author}{S.~Chobsaard},
\newblock \bibinfo{title}{The relationship between spectral correlation and
  envelope analysis in the diagnostics of bearing faults and other
  cyclostationary machine signals},
\newblock \bibinfo{journal}{Mechanical Systems and Signal Processing}
  \bibinfo{volume}{15} (\bibinfo{year}{2001}) \bibinfo{pages}{945--962}.
\bibitem[{McFadden and Smith(1984)}]{mcfadden1984model}
\bibinfo{author}{P.~McFadden}, \bibinfo{author}{J.~Smith},
\newblock \bibinfo{title}{Model for the vibration produced by a single point
  defect in a rolling element bearing},
\newblock \bibinfo{journal}{Journal of Sound and Vibration}
  \bibinfo{volume}{96} (\bibinfo{year}{1984}) \bibinfo{pages}{69--82}.
\bibitem[{Feng et~al.(2013)Feng, Liang, and Chu}]{feng2013recent}
\bibinfo{author}{Z.~Feng}, \bibinfo{author}{M.~Liang},
  \bibinfo{author}{F.~Chu},
\newblock \bibinfo{title}{Recent advances in time--frequency analysis methods
  for machinery fault diagnosis: A review with application examples},
\newblock \bibinfo{journal}{Mechanical Systems and Signal Processing}
  \bibinfo{volume}{38} (\bibinfo{year}{2013}) \bibinfo{pages}{165--205}.
\bibitem[{Stankovic(1994)}]{stankovic1994method}
\bibinfo{author}{L.~Stankovic},
\newblock \bibinfo{title}{A method for time-frequency analysis},
\newblock \bibinfo{journal}{IEEE Transactions on Signal Processing}
  \bibinfo{volume}{42} (\bibinfo{year}{1994}) \bibinfo{pages}{225--229}.
\bibitem[{Yu(2019)}]{yu2019concentrated}
\bibinfo{author}{G.~Yu},
\newblock \bibinfo{title}{A concentrated time--frequency analysis tool for
  bearing fault diagnosis},
\newblock \bibinfo{journal}{IEEE Transactions on Instrumentation and
  Measurement} \bibinfo{volume}{69} (\bibinfo{year}{2019})
  \bibinfo{pages}{371--381}.
\bibitem[{Napolitano(2016)}]{napolitano2016cyclostationarity}
\bibinfo{author}{A.~Napolitano},
\newblock \bibinfo{title}{Cyclostationarity: New trends and applications},
\newblock \bibinfo{journal}{Signal Processing} \bibinfo{volume}{120}
  (\bibinfo{year}{2016}) \bibinfo{pages}{385--408}.
\bibitem[{Lou and Loparo(2004)}]{lou2004bearing}
\bibinfo{author}{X.~Lou}, \bibinfo{author}{K.~A. Loparo},
\newblock \bibinfo{title}{Bearing fault diagnosis based on wavelet transform
  and fuzzy inference},
\newblock \bibinfo{journal}{Mechanical Systems and Signal Processing}
  \bibinfo{volume}{18} (\bibinfo{year}{2004}) \bibinfo{pages}{1077--1095}.
\bibitem[{Gao et~al.(2015)Gao, Liang, Chen, and Xu}]{gao2015feature}
\bibinfo{author}{H.~Gao}, \bibinfo{author}{L.~Liang},
  \bibinfo{author}{X.~Chen}, \bibinfo{author}{G.~Xu},
\newblock \bibinfo{title}{Feature extraction and recognition for rolling
  element bearing fault utilizing short-time fourier transform and non-negative
  matrix factorization},
\newblock \bibinfo{journal}{Chinese Journal of Mechanical Engineering}
  \bibinfo{volume}{28} (\bibinfo{year}{2015}) \bibinfo{pages}{96--105}.
\bibitem[{Feng et~al.(2017)Feng, Zhang, and
  Zuo}]{fengAdaptiveModeDecomposition2017}
\bibinfo{author}{Z.~Feng}, \bibinfo{author}{D.~Zhang}, \bibinfo{author}{M.~J.
  Zuo},
\newblock \bibinfo{title}{Adaptive mode decomposition methods and their
  applications in signal analysis for machinery fault diagnosis: A review with
  examples},
\newblock \bibinfo{journal}{IEEE Access} \bibinfo{volume}{5}
  (\bibinfo{year}{2017}) \bibinfo{pages}{24301--24331}.
\bibitem[{Azar et~al.(2022)Azar, Hajiakhondi-Meybodi, and
  Naderkhani}]{azar2022semi}
\bibinfo{author}{K.~Azar}, \bibinfo{author}{Z.~Hajiakhondi-Meybodi},
  \bibinfo{author}{F.~Naderkhani},
\newblock \bibinfo{title}{Semi-supervised clustering-based method for fault
  diagnosis and prognosis: A case study},
\newblock \bibinfo{journal}{Reliability Engineering \& System Safety}
  \bibinfo{volume}{222} (\bibinfo{year}{2022}) \bibinfo{pages}{108405}.
\bibitem[{Nemani et~al.(2023)Nemani, Biggio, Huan, Hu, Fink, Tran, Wang, Du,
  Zhang, and Hu}]{nemani2023uncertainty}
\bibinfo{author}{V.~Nemani}, \bibinfo{author}{L.~Biggio},
  \bibinfo{author}{X.~Huan}, \bibinfo{author}{Z.~Hu},
  \bibinfo{author}{O.~Fink}, \bibinfo{author}{A.~Tran},
  \bibinfo{author}{Y.~Wang}, \bibinfo{author}{X.~Du},
  \bibinfo{author}{X.~Zhang}, \bibinfo{author}{C.~Hu},
\newblock \bibinfo{title}{Uncertainty quantification in machine learning for
  engineering design and health prognostics: A tutorial},
\newblock \bibinfo{journal}{arXiv preprint arXiv:2305.04933}
  (\bibinfo{year}{2023}).
\bibitem[{Zhang et~al.(2017)Zhang, Peng, Li, Chen, and Zhang}]{zhang2017new}
\bibinfo{author}{W.~Zhang}, \bibinfo{author}{G.~Peng}, \bibinfo{author}{C.~Li},
  \bibinfo{author}{Y.~Chen}, \bibinfo{author}{Z.~Zhang},
\newblock \bibinfo{title}{A new deep learning model for fault diagnosis with
  good anti-noise and domain adaptation ability on raw vibration signals},
\newblock \bibinfo{journal}{Sensors} \bibinfo{volume}{17}
  (\bibinfo{year}{2017}) \bibinfo{pages}{425}.
\bibitem[{Zhao et~al.(2019)Zhao, Zhong, Fu, Tang, and Pecht}]{zhao2019deep}
\bibinfo{author}{M.~Zhao}, \bibinfo{author}{S.~Zhong}, \bibinfo{author}{X.~Fu},
  \bibinfo{author}{B.~Tang}, \bibinfo{author}{M.~Pecht},
\newblock \bibinfo{title}{Deep residual shrinkage networks for fault
  diagnosis},
\newblock \bibinfo{journal}{IEEE Transactions on Industrial Informatics}
  \bibinfo{volume}{16} (\bibinfo{year}{2019}) \bibinfo{pages}{4681--4690}.
\bibitem[{Liu et~al.(2016)Liu, Meng, Yang, Sun, and Chen}]{liu2016dislocated}
\bibinfo{author}{R.~Liu}, \bibinfo{author}{G.~Meng}, \bibinfo{author}{B.~Yang},
  \bibinfo{author}{C.~Sun}, \bibinfo{author}{X.~Chen},
\newblock \bibinfo{title}{Dislocated time series convolutional neural
  architecture: An intelligent fault diagnosis approach for electric machine},
\newblock \bibinfo{journal}{IEEE Transactions on Industrial Informatics}
  \bibinfo{volume}{13} (\bibinfo{year}{2016}) \bibinfo{pages}{1310--1320}.
\bibitem[{Zuo et~al.(2022)Zuo, Xu, Zhang, Xiahou, and Liu}]{zuo2022multi}
\bibinfo{author}{L.~Zuo}, \bibinfo{author}{F.~Xu}, \bibinfo{author}{C.~Zhang},
  \bibinfo{author}{T.~Xiahou}, \bibinfo{author}{Y.~Liu},
\newblock \bibinfo{title}{A multi-layer spiking neural network-based approach
  to bearing fault diagnosis},
\newblock \bibinfo{journal}{Reliability Engineering \& System Safety}
  \bibinfo{volume}{225} (\bibinfo{year}{2022}) \bibinfo{pages}{108561}.
\bibitem[{Yang et~al.(2019)Yang, Lei, Jia, and Xing}]{yang2019intelligent}
\bibinfo{author}{B.~Yang}, \bibinfo{author}{Y.~Lei}, \bibinfo{author}{F.~Jia},
  \bibinfo{author}{S.~Xing},
\newblock \bibinfo{title}{An intelligent fault diagnosis approach based on
  transfer learning from laboratory bearings to locomotive bearings},
\newblock \bibinfo{journal}{Mechanical Systems and Signal Processing}
  \bibinfo{volume}{122} (\bibinfo{year}{2019}) \bibinfo{pages}{692--706}.
\bibitem[{Chen et~al.(2022)Chen, Chen, Feng, Liu, Zhang, Zhang, and
  Xiao}]{chen2022imbalance}
\bibinfo{author}{Z.~Chen}, \bibinfo{author}{J.~Chen},
  \bibinfo{author}{Y.~Feng}, \bibinfo{author}{S.~Liu},
  \bibinfo{author}{T.~Zhang}, \bibinfo{author}{K.~Zhang},
  \bibinfo{author}{W.~Xiao},
\newblock \bibinfo{title}{Imbalance fault diagnosis under long-tailed
  distribution: Challenges, solutions and prospects},
\newblock \bibinfo{journal}{Knowledge-Based Systems} \bibinfo{volume}{258}
  (\bibinfo{year}{2022}) \bibinfo{pages}{110008}.
\bibitem[{Cheng et~al.(2023)Cheng, Xie, Xing, Nie, Chen, Liu, Liu, Huang, and
  Zhang}]{10158933}
\bibinfo{author}{W.~Cheng}, \bibinfo{author}{S.~Xie},
  \bibinfo{author}{J.~Xing}, \bibinfo{author}{Z.~Nie},
  \bibinfo{author}{X.~Chen}, \bibinfo{author}{Y.~Liu},
  \bibinfo{author}{X.~Liu}, \bibinfo{author}{Q.~Huang},
  \bibinfo{author}{R.~Zhang},
\newblock \bibinfo{title}{Interactive hybrid model for remaining useful life
  prediction with uncertainty quantification of bearing in nuclear circulating
  water pump},
\newblock \bibinfo{journal}{IEEE Transactions on Industrial Informatics}
  (\bibinfo{year}{2023}) \bibinfo{pages}{1--12}.
\bibitem[{Harunuzzaman and Aldemir(1996)}]{harunuzzaman1996optimization}
\bibinfo{author}{M.~Harunuzzaman}, \bibinfo{author}{T.~Aldemir},
\newblock \bibinfo{title}{Optimization of standby safety system maintenance
  schedules in nuclear power plants},
\newblock \bibinfo{journal}{Nuclear Technology} \bibinfo{volume}{113}
  (\bibinfo{year}{1996}) \bibinfo{pages}{354--367}.
\bibitem[{Zhang et~al.(2021)Zhang, Kang, Hooi, Yan, and Feng}]{zhang2021deep}
\bibinfo{author}{Y.~Zhang}, \bibinfo{author}{B.~Kang},
  \bibinfo{author}{B.~Hooi}, \bibinfo{author}{S.~Yan},
  \bibinfo{author}{J.~Feng},
\newblock \bibinfo{title}{Deep long-tailed learning: A survey},
\newblock \bibinfo{journal}{arXiv preprint arXiv:2110.04596}
  (\bibinfo{year}{2021}).
\bibitem[{Yang et~al.(2020)Yang, Xie, and Yang}]{yang2020improved}
\bibinfo{author}{J.~Yang}, \bibinfo{author}{G.~Xie}, \bibinfo{author}{Y.~Yang},
\newblock \bibinfo{title}{An improved ensemble fusion autoencoder model for
  fault diagnosis from imbalanced and incomplete data},
\newblock \bibinfo{journal}{Control Engineering Practice} \bibinfo{volume}{98}
  (\bibinfo{year}{2020}) \bibinfo{pages}{104358}.
\bibitem[{Khoshgoftaar and Gao(2009)}]{khoshgoftaar2009feature}
\bibinfo{author}{T.~M. Khoshgoftaar}, \bibinfo{author}{K.~Gao},
\newblock \bibinfo{title}{Feature selection with imbalanced data for software
  defect prediction},
\newblock in: \bibinfo{booktitle}{2009 International Conference on Machine
  Learning and Applications}, \bibinfo{year}{2009}, pp.
  \bibinfo{pages}{235--240}. \DOIprefix\doi{10.1109/ICMLA.2009.18}.
\bibitem[{Liu et~al.(2019)Liu, Ma, and Cheng}]{liu2019generative}
\bibinfo{author}{Q.~Liu}, \bibinfo{author}{G.~Ma}, \bibinfo{author}{C.~Cheng},
\newblock \bibinfo{title}{Generative adversarial network based multi-class
  imbalanced fault diagnosis of rolling bearing},
\newblock in: \bibinfo{booktitle}{2019 4th International Conference on System
  Reliability and Safety (ICSRS)}, \bibinfo{organization}{IEEE},
  \bibinfo{year}{2019}, pp. \bibinfo{pages}{318--324}.
\bibitem[{Li et~al.(2021)Li, Jiang, Liu, Zhang, and Xu}]{li2021unified}
\bibinfo{author}{X.~Li}, \bibinfo{author}{H.~Jiang}, \bibinfo{author}{S.~Liu},
  \bibinfo{author}{J.~Zhang}, \bibinfo{author}{J.~Xu},
\newblock \bibinfo{title}{A unified framework incorporating predictive
  generative denoising autoencoder and deep coral network for rolling bearing
  fault diagnosis with unbalanced data},
\newblock \bibinfo{journal}{Measurement} \bibinfo{volume}{178}
  (\bibinfo{year}{2021}) \bibinfo{pages}{109345}.
\bibitem[{Zhang et~al.(2022)Zhang, Zou, Su, Tang, Kang, Xu, Liu, and
  Fan}]{zhang2022class}
\bibinfo{author}{J.~Zhang}, \bibinfo{author}{J.~Zou}, \bibinfo{author}{Z.~Su},
  \bibinfo{author}{J.~Tang}, \bibinfo{author}{Y.~Kang},
  \bibinfo{author}{H.~Xu}, \bibinfo{author}{Z.~Liu}, \bibinfo{author}{S.~Fan},
\newblock \bibinfo{title}{A class-aware supervised contrastive learning
  framework for imbalanced fault diagnosis},
\newblock \bibinfo{journal}{Knowledge-Based Systems} \bibinfo{volume}{252}
  (\bibinfo{year}{2022}) \bibinfo{pages}{109437}.
\bibitem[{Peng et~al.(2022)Peng, Lu, Tao, Ma, Zhang, Wang, and
  Zhang}]{peng2022progressively}
\bibinfo{author}{P.~Peng}, \bibinfo{author}{J.~Lu}, \bibinfo{author}{S.~Tao},
  \bibinfo{author}{K.~Ma}, \bibinfo{author}{Y.~Zhang},
  \bibinfo{author}{H.~Wang}, \bibinfo{author}{H.~Zhang},
\newblock \bibinfo{title}{Progressively balanced supervised contrastive
  representation learning for long-tailed fault diagnosis},
\newblock \bibinfo{journal}{IEEE Transactions on Instrumentation and
  Measurement} \bibinfo{volume}{71} (\bibinfo{year}{2022})
  \bibinfo{pages}{1--12}.
\bibitem[{Hou et~al.(2022)Hou, Chen, Feng, Liu, He, and
  Zhou}]{hou2022contrastive}
\bibinfo{author}{R.~Hou}, \bibinfo{author}{J.~Chen}, \bibinfo{author}{Y.~Feng},
  \bibinfo{author}{S.~Liu}, \bibinfo{author}{S.~He}, \bibinfo{author}{Z.~Zhou},
\newblock \bibinfo{title}{Contrastive-weighted self-supervised model for
  long-tailed data classification with vision transformer augmented},
\newblock \bibinfo{journal}{Mechanical Systems and Signal Processing}
  \bibinfo{volume}{177} (\bibinfo{year}{2022}) \bibinfo{pages}{109174}.
\bibitem[{Khosla et~al.(2020)Khosla, Teterwak, Wang, Sarna, Tian, Isola,
  Maschinot, Liu, and Krishnan}]{khosla2020supervised}
\bibinfo{author}{P.~Khosla}, \bibinfo{author}{P.~Teterwak},
  \bibinfo{author}{C.~Wang}, \bibinfo{author}{A.~Sarna},
  \bibinfo{author}{Y.~Tian}, \bibinfo{author}{P.~Isola},
  \bibinfo{author}{A.~Maschinot}, \bibinfo{author}{C.~Liu},
  \bibinfo{author}{D.~Krishnan},
\newblock \bibinfo{title}{Supervised contrastive learning},
\newblock \bibinfo{journal}{Advances in Neural Information Processing Systems}
  \bibinfo{volume}{33} (\bibinfo{year}{2020}) \bibinfo{pages}{18661--18673}.
\bibitem[{Li et~al.(2023)Li, Lei, Huang, Nazeer, Long, and Yang}]{10042974}
\bibinfo{author}{C.~Li}, \bibinfo{author}{X.~Lei}, \bibinfo{author}{Y.~Huang},
  \bibinfo{author}{F.~Nazeer}, \bibinfo{author}{J.~Long},
  \bibinfo{author}{Z.~Yang},
\newblock \bibinfo{title}{Incrementally contrastive learning of homologous and
  interclass features for the fault diagnosis of rolling element bearings},
\newblock \bibinfo{journal}{IEEE Transactions on Industrial Informatics}
  (\bibinfo{year}{2023}) \bibinfo{pages}{1--9}.
\bibitem[{Zhang et~al.(2023)Zhang, Chen, Xiao, and Yin}]{10018491}
\bibinfo{author}{W.~Zhang}, \bibinfo{author}{D.~Chen},
  \bibinfo{author}{Y.~Xiao}, \bibinfo{author}{H.~Yin},
\newblock \bibinfo{title}{Semi-supervised contrast learning based on
  multi-scale attention and multi-target contrast learning for bearing fault
  diagnosis},
\newblock \bibinfo{journal}{IEEE Transactions on Industrial Informatics}
  (\bibinfo{year}{2023}) \bibinfo{pages}{1--13}.
\bibitem[{Fan et~al.(2023)Fan, Dong, Wu, Ruan, Zeng, Cui, and
  Liao}]{fan2023one}
\bibinfo{author}{F.-L. Fan}, \bibinfo{author}{H.-C. Dong},
  \bibinfo{author}{Z.~Wu}, \bibinfo{author}{L.~Ruan},
  \bibinfo{author}{T.~Zeng}, \bibinfo{author}{Y.~Cui}, \bibinfo{author}{J.-X.
  Liao},
\newblock \bibinfo{title}{One neuron saved is one neuron earned: On parametric
  efficiency of quadratic networks},
\newblock \bibinfo{journal}{arXiv preprint arXiv:2303.06316}
  (\bibinfo{year}{2023}).
\bibitem[{Liao et~al.(2023)Liao, Dong, Sun, Sun, Zhang, and
  Fan}]{liao2023attention}
\bibinfo{author}{J.-X. Liao}, \bibinfo{author}{H.-C. Dong},
  \bibinfo{author}{Z.-Q. Sun}, \bibinfo{author}{J.~Sun},
  \bibinfo{author}{S.~Zhang}, \bibinfo{author}{F.-L. Fan},
\newblock \bibinfo{title}{Attention-embedded quadratic network (qttention) for
  effective and interpretable bearing fault diagnosis},
\newblock \bibinfo{journal}{IEEE Transactions on Instrumentation and
  Measurement}  (\bibinfo{year}{2023}).
\bibitem[{Graf et~al.(2021)Graf, Hofer, Niethammer, and
  Kwitt}]{graf2021dissecting}
\bibinfo{author}{F.~Graf}, \bibinfo{author}{C.~Hofer},
  \bibinfo{author}{M.~Niethammer}, \bibinfo{author}{R.~Kwitt},
\newblock \bibinfo{title}{Dissecting supervised contrastive learning},
\newblock in: \bibinfo{booktitle}{International Conference on Machine
  Learning}, \bibinfo{organization}{PMLR}, \bibinfo{year}{2021}, pp.
  \bibinfo{pages}{3821--3830}.
\bibitem[{Wu et~al.(2021)Wu, Zhao, Sun, Yan, and Chen}]{wu2021learning}
\bibinfo{author}{J.~Wu}, \bibinfo{author}{Z.~Zhao}, \bibinfo{author}{C.~Sun},
  \bibinfo{author}{R.~Yan}, \bibinfo{author}{X.~Chen},
\newblock \bibinfo{title}{Learning from class-imbalanced data with a
  model-agnostic framework for machine intelligent diagnosis},
\newblock \bibinfo{journal}{Reliability Engineering \& System Safety}
  \bibinfo{volume}{216} (\bibinfo{year}{2021}) \bibinfo{pages}{107934}.
\bibitem[{Han et~al.(2022)Han, Shao, Huo, Yang, and Cheng}]{han2022end}
\bibinfo{author}{S.~Han}, \bibinfo{author}{H.~Shao}, \bibinfo{author}{Z.~Huo},
  \bibinfo{author}{X.~Yang}, \bibinfo{author}{J.~Cheng},
\newblock \bibinfo{title}{End-to-end chiller fault diagnosis using fused
  attention mechanism and dynamic cross-entropy under imbalanced datasets},
\newblock \bibinfo{journal}{Building and Environment} \bibinfo{volume}{212}
  (\bibinfo{year}{2022}) \bibinfo{pages}{108821}.
\bibitem[{Wang and Liu(2021)}]{wang2021understanding}
\bibinfo{author}{F.~Wang}, \bibinfo{author}{H.~Liu},
\newblock \bibinfo{title}{Understanding the behaviour of contrastive loss},
\newblock in: \bibinfo{booktitle}{Proceedings of the IEEE/CVF Conference on
  Computer Vision and Pattern Recognition}, \bibinfo{year}{2021}, pp.
  \bibinfo{pages}{2495--2504}.
\bibitem[{Wang et~al.(2021)Wang, Han, Wei, Zhang, and
  Wang}]{wang2021contrastive}
\bibinfo{author}{P.~Wang}, \bibinfo{author}{K.~Han}, \bibinfo{author}{X.-S.
  Wei}, \bibinfo{author}{L.~Zhang}, \bibinfo{author}{L.~Wang},
\newblock \bibinfo{title}{Contrastive learning based hybrid networks for
  long-tailed image classification},
\newblock in: \bibinfo{booktitle}{Proceedings of the IEEE/CVF conference on
  computer vision and pattern recognition}, \bibinfo{year}{2021}, pp.
  \bibinfo{pages}{943--952}.
\bibitem[{Yang et~al.(2022)Yang, Wang, and Zhu}]{yang2022few}
\bibinfo{author}{Z.~Yang}, \bibinfo{author}{J.~Wang}, \bibinfo{author}{Y.~Zhu},
\newblock \bibinfo{title}{Few-shot classification with contrastive learning},
\newblock in: \bibinfo{booktitle}{Computer Vision--ECCV 2022: 17th European
  Conference, Tel Aviv, Israel, October 23--27, 2022, Proceedings, Part XX},
  \bibinfo{organization}{Springer}, \bibinfo{year}{2022}, pp.
  \bibinfo{pages}{293--309}.
\bibitem[{Hu et~al.(2021)Hu, Wu, Sun, Yan, and Chen}]{hu2021robust}
\bibinfo{author}{C.~Hu}, \bibinfo{author}{J.~Wu}, \bibinfo{author}{C.~Sun},
  \bibinfo{author}{R.~Yan}, \bibinfo{author}{X.~Chen},
\newblock \bibinfo{title}{Robust supervised contrastive learning for fault
  diagnosis under different noises and conditions},
\newblock in: \bibinfo{booktitle}{2021 International Conference on Sensing,
  Measurement \& Data Analytics in the era of Artificial Intelligence (ICSMD)},
  \bibinfo{organization}{IEEE}, \bibinfo{year}{2021}, pp.
  \bibinfo{pages}{1--6}.
\bibitem[{Fan et~al.(2019)Fan, Shan, Kalra, Singh, Qian, Getzin, Teng, Hahn,
  and Wang}]{fan2019quadratic}
\bibinfo{author}{F.~Fan}, \bibinfo{author}{H.~Shan}, \bibinfo{author}{M.~K.
  Kalra}, \bibinfo{author}{R.~Singh}, \bibinfo{author}{G.~Qian},
  \bibinfo{author}{M.~Getzin}, \bibinfo{author}{Y.~Teng},
  \bibinfo{author}{J.~Hahn}, \bibinfo{author}{G.~Wang},
\newblock \bibinfo{title}{Quadratic autoencoder (q-ae) for low-dose ct
  denoising},
\newblock \bibinfo{journal}{IEEE Transactions on Medical Imaging}
  \bibinfo{volume}{39} (\bibinfo{year}{2019}) \bibinfo{pages}{2035--2050}.
\bibitem[{Fan et~al.(2021)Fan, Li, Wang, Lai, and Wang}]{fan2021expressivity}
\bibinfo{author}{F.-L. Fan}, \bibinfo{author}{M.~Li},
  \bibinfo{author}{F.~Wang}, \bibinfo{author}{R.~Lai},
  \bibinfo{author}{G.~Wang},
\newblock \bibinfo{title}{Expressivity and trainability of quadratic networks},
\newblock \bibinfo{journal}{arXiv preprint arXiv:2110.06081}
  (\bibinfo{year}{2021}).
\bibitem[{Fan et~al.(2020)Fan, Xiong, and Wang}]{fan2020universal}
\bibinfo{author}{F.~Fan}, \bibinfo{author}{J.~Xiong},
  \bibinfo{author}{G.~Wang},
\newblock \bibinfo{title}{Universal approximation with quadratic deep
  networks},
\newblock \bibinfo{journal}{Neural Networks} \bibinfo{volume}{124}
  (\bibinfo{year}{2020}) \bibinfo{pages}{383--392}.
\bibitem[{He et~al.(2015)He, Zhang, Ren, and Sun}]{he2015delving}
\bibinfo{author}{K.~He}, \bibinfo{author}{X.~Zhang}, \bibinfo{author}{S.~Ren},
  \bibinfo{author}{J.~Sun},
\newblock \bibinfo{title}{Delving deep into rectifiers: Surpassing human-level
  performance on imagenet classification},
\newblock in: \bibinfo{booktitle}{Proceedings of the IEEE International
  Conference on Computer Vision}, \bibinfo{year}{2015}, pp.
  \bibinfo{pages}{1026--1034}.
\bibitem[{Zhu et~al.(2022)Zhu, Wang, Chen, Chen, and Jiang}]{zhu2022balanced}
\bibinfo{author}{J.~Zhu}, \bibinfo{author}{Z.~Wang}, \bibinfo{author}{J.~Chen},
  \bibinfo{author}{Y.-P.~P. Chen}, \bibinfo{author}{Y.-G. Jiang},
\newblock \bibinfo{title}{Balanced contrastive learning for long-tailed visual
  recognition},
\newblock in: \bibinfo{booktitle}{Proceedings of the IEEE/CVF Conference on
  Computer Vision and Pattern Recognition}, \bibinfo{year}{2022}, pp.
  \bibinfo{pages}{6908--6917}.
\bibitem[{Menon et~al.(2020)Menon, Jayasumana, Rawat, Jain, Veit, and
  Kumar}]{menon2020long}
\bibinfo{author}{A.~K. Menon}, \bibinfo{author}{S.~Jayasumana},
  \bibinfo{author}{A.~S. Rawat}, \bibinfo{author}{H.~Jain},
  \bibinfo{author}{A.~Veit}, \bibinfo{author}{S.~Kumar},
\newblock \bibinfo{title}{Long-tail learning via logit adjustment},
\newblock \bibinfo{journal}{arXiv preprint arXiv:2007.07314}
  (\bibinfo{year}{2020}).
\bibitem[{Chen et~al.(2022)Chen, Yang, and Liu}]{chen2022self}
\bibinfo{author}{J.~Chen}, \bibinfo{author}{B.~Yang}, \bibinfo{author}{R.~Liu},
\newblock \bibinfo{title}{Self-supervised contrastive learning approach for
  bearing fault diagnosis with rare labeled data},
\newblock in: \bibinfo{booktitle}{2022 IEEE 31st International Symposium on
  Industrial Electronics (ISIE)}, \bibinfo{organization}{IEEE},
  \bibinfo{year}{2022}, pp. \bibinfo{pages}{1190--1194}.
\bibitem[{Jia et~al.(2018)Jia, Lei, Lu, and Xing}]{jia2018deep}
\bibinfo{author}{F.~Jia}, \bibinfo{author}{Y.~Lei}, \bibinfo{author}{N.~Lu},
  \bibinfo{author}{S.~Xing},
\newblock \bibinfo{title}{Deep normalized convolutional neural network for
  imbalanced fault classification of machinery and its understanding via
  visualization},
\newblock \bibinfo{journal}{Mechanical Systems and Signal Processing}
  \bibinfo{volume}{110} (\bibinfo{year}{2018}) \bibinfo{pages}{349--367}.

\end{thebibliography}

\end{document}